\documentclass[twocolumn]{aastex631}
\usepackage{amsmath}
\usepackage{url}

\shorttitle{Splashback Mass Function from Photometry}
\shortauthors{Gabriel-Silva \& Sodr\'{e}}

\begin{document}

\title{The Splashback Mass Function of Galaxy Clusters from Photometric Data}

\author[0000-0002-8881-0694]{Lucas Gabriel-Silva}
\affiliation{Instituto de Astronomia, Geof\'{i}sica e Ci\^{e}ncias Atmosf\'{e}ricas da Universidade de S\~{a}o Paulo (IAG/USP), \\
Rua do Mat\~{a}o 1226, Cidade Universit\'{a}ria, S\~{a}o Paulo, 05508-090, Brazil}
\email{lucasgabriel@usp.br}

\author[0000-0002-3876-268X]{Laerte Sodr\'{e} Jr.}
\affiliation{Instituto de Astronomia, Geof\'{i}sica e Ci\^{e}ncias Atmosf\'{e}ricas da Universidade de S\~{a}o Paulo (IAG/USP), \\
Rua do Mat\~{a}o 1226, Cidade Universit\'{a}ria, S\~{a}o Paulo, 05508-090, Brazil}
\email{laerte.sodre@iag.usp.br}

\begin{abstract}
The splashback radius marks the physical boundary of galaxy clusters, separating orbiting from infalling material, and provides a halo definition free from pseudo-evolution. In this work, we present a fully photometric framework to measure individual cluster splashback radii and masses, and to construct an observational splashback mass function. Using Sloan Digital Sky Survey data, we develop a probabilistic cluster membership method based on radial and photometric redshift information, optimized through an adaptive probability cut that maximizes the detection significance of the cluster core relative to its outskirts. We apply this methodology to a sample of 499 galaxy clusters from the \textsc{CoMaLit} weak-lensing compilation and recover splashback radii from modeling cumulative galaxy number profiles. The resulting splashback radii exhibit a median ratio $R_{\mathrm{sp}}/R_{200\mathrm{m}} \simeq 1.1$, consistent with previous observational studies. Using these measurements, we recalibrate the $M_{\mathrm{sp}}$--$R_{\mathrm{sp}}$ scaling relation over a wide redshift range ($0.01 < z < 0.8$), finding a slope shallower than the constant-density expectation and no significant redshift evolution. We then apply this relation to \textsc{redMaPPer} clusters in the SDSS Northern Galactic Cap to derive splashback masses for more than $1.5\times10^4$ systems and construct the first observational splashback mass function based solely on photometric data. The resulting mass function agrees with simulation-based predictions at the high-mass end, while deviations at lower masses are consistent with known completeness limits of optical cluster catalogs. Our results demonstrate that splashback-based cluster sizes, masses, and abundances can be robustly measured in photometric surveys, enabling cosmological studies without spectroscopic or lensing data.
\end{abstract}


\section{Introduction}\label{sec:intro}

Although conceptually simple, the number of dark matter halos as a function of their mass is one of the most important statistics in cosmology. The halo mass function establishes a fundamental link between theoretical predictions of structure formation and observations, particularly when galaxy cluster abundances are used to constrain cosmological parameters \citep[e.g.,][]{vikhlinin+09, allen+11, kravtsov+12}. 

Early theoretical descriptions of this connection date back to the pioneering work of \citet{press+74}, who proposed that gravitational collapse occurs at the locations of sufficiently high peaks in an initially Gaussian density field. Within this framework, halo formation is governed by a critical overdensity threshold, leading to a mass function that is assumed to be universal, with little explicit dependence on redshift or cosmology. While later developments have revealed important deviations from this idealized picture \citep[e.g.,][]{sheth+01, tinker+08}, this early model already emphasized the key role of galaxy clusters as tracers of structure growth and the importance of robust mass definitions for cosmological applications.

However, robust estimates of galaxy cluster masses usually rely on observationally expensive or methodologically complex techniques. These include spectroscopic measurements based on the virial theorem \citep[e.g.,][]{carlberg+97}, X-ray observations of the intra-cluster medium used to infer the underlying gravitational potential \citep[e.g.,][]{arnaud+05}, and weak gravitational lensing analyses \citep[e.g.,][]{umetsu+20}. At the same time, the advent of large photometric surveys such as the Southern Photometric Local Universe Survey (S-PLUS; \citealt{mendes+19}), the Javalambre Photometric Local Universe Survey (J-PLUS; \citealt{cenarro+19}), and the Javalambre-Physics of the Accelerating Universe Astrophysical Survey (J-PAS; \citealt{bonoli+21}) motivates the development of alternative methods capable of estimating cluster masses using solely photometric data.

Traditionally, the characteristic mass and size of galaxy clusters, and of dark matter halos in general, have been defined through spherical overdensity criteria. In this framework, the halo mass is defined within a radius enclosing a fixed overdensity relative to a reference density, such as the critical (c) or mean (m) density of the Universe\footnote{The critical density is defined as $\rho_{\mathrm{c}}(z)=3H^2(z)/(8\pi G)$ and corresponds to the density required for a spatially flat Universe at redshift $z$. The mean matter density is given by $\rho_{\mathrm{m}}(z)=\Omega_{\mathrm{M}}(z)\,\rho_{\mathrm{c}}(z)$ and represents the average matter density of the Universe at the same redshift.}. Common examples include $M_{200c}$, $M_{200m}$, and $M_{\rm vir}$, with corresponding radii $R_{200c}$, $R_{200m}$, and $R_{\rm vir}$ \citep[e.g.,][]{gunn+72, lacey+93}. While widely used, these definitions present several limitations. For instance, satellite galaxies may orbit well beyond the virial radius \citep[e.g.,][]{wetzel+14}, and infalling subhalos can experience tidal stripping at radii significantly larger than $R_{\rm vir}$ \citep{behroozi+14}. Moreover, as demonstrated by \citet{diemer+13}, spherical overdensity boundaries are affected by pseudo-evolution: as the reference density decreases with cosmic expansion, halo masses and radii may appear to grow even if the physical density profile remains unchanged.

A more physically motivated halo boundary is provided by the splashback radius, $R_{\rm sp}$, which corresponds to the apocenter of recently accreted material orbiting within the cluster potential \citep[e.g.,][]{diemer+14, adhikari+14, more+15, diemer22}. This radius marks the transition between virialized and infalling matter and manifests observationally as a steepening in the density profile. Such splashback features have now been detected in numerous studies \citep[e.g.,][]{more+16, baxter+17, adhikari+21, xu+24, gabriel-silva+25}. Owing to its dynamical origin, $R_{\rm sp}$ is expected to encode information about halo assembly history and mass accretion rates, providing a direct link between cluster outskirts and cosmological growth \citep[e.g.,][]{more+15, dacunha+25, joshi+25, walker+25}.

Beyond its location, the splashback feature itself has been shown to contain additional information about halo structure and evolution. Using high-resolution hydrodynamic simulations from the MillenniumTNG project, \citet{yu+25} introduced quantitative descriptors of the splashback feature, namely its depth and width. They demonstrated that these quantities correlate strongly with long-term halo properties such as mass, peak height, concentration, and formation time, while exhibiting only weak sensitivity to short-term processes like recent mergers or instantaneous mass accretion. These results suggest that the shape of the splashback feature acts as a long-term memory of halo growth, offering a complementary and physically motivated characterisation of halo boundaries beyond traditional spherical overdensity definitions.

Observationally, splashback features have been detected using satellite galaxy distributions in optically selected galaxy clusters \citep[e.g.,][]{more+16, baxter+17, gabriel-silva+25}, in clusters selected via the Sunyaev--Zel'dovich effect (SZ) \citep[e.g.,][]{shin+19, zurcher+19, adhikari+21}, and using weak gravitational lensing measurements (WL) \citep{chang+18, contigiani+19, xu+24, joshi+25}. More recently, splashback-like signatures have also been identified in the diffuse intra-cluster light (ICL), which originates from the debris of disrupted galaxies and serves as a powerful tracer of cluster outskirts \citep{montes+19, contini21}. Using simulations from The Three Hundred Project, \citet{walker+25} showed that the steepening in the stellar density profile is consistent with that of the dark matter, albeit systematically sharper. They further demonstrated that the relation between the splashback feature and the cluster mass accretion rate is consistent with expectations from dark-matter-only simulations.

Evidence for splashback-like features has also emerged on smaller scales. \citet{zhang+25} reported the detection of a splashback-like signal in the outer regions of central galaxies within clusters, observed as a dip in the radial slope of the surface brightness profile. This signal was identified through stacking Dark Energy Survey data for more than four thousand galaxy clusters in the redshift range $0.2 < z < 0.5$. The location of the dip, typically between 40 and 60 kpc from the galaxy centre, resembles the transition associated with the splashback effect, and comparison with simulations suggests that it may mark the boundary between the central galaxy and the surrounding diffuse cluster light undergoing recent or ongoing accretion.

Given the physical robustness of $R_{\rm sp}$, it is natural to define an associated mass, the splashback mass ($M_{\rm sp}$), which, like the radius, is free from pseudo-evolution. The splashback mass provides a physically motivated estimate of the total matter accreted by a halo by a given epoch, making it a promising tracer of structure growth over cosmic time and a potentially powerful tool for cosmological analyses. Nevertheless, splashback-based mass definitions are still relatively underexplored and not yet widely adopted in observational studies \citep[e.g.,][]{ryu+21, diemer20}.

In this work, we extend the results of \citet{gabriel-silva+25}, who demonstrated the feasibility of estimating splashback masses from splashback radii using spectroscopic data by a fitted $R_{\mathrm{sp}}$--$M_{\mathrm{sp}}$ relation. Here, we refine and better constrain this relation, and extend it to photometric data, taking advantage of the large volume and depth of the Sloan Digital Sky Survey. We show that it is possible to recover the splashback radius using only photometric redshifts and, furthermore, to reconstruct galaxy cluster mass functions based on these splashback-based estimates.

This paper is organised as follows. In Section~\ref{sec:data}, we describe the observational data sets; in Section~\ref{sec:methods}, we present the probabilistic membership assignment and cumulative profile modelling; in Section~\ref{sec:results}, we discuss our results and compare them with previous studies; and in Section~\ref{sec:conclusion}, we summarise our findings and outline future prospects. Throughout the work, we assume a flat $\Lambda$CDM cosmology with $\Omega_M = 0.28$, $\Omega_\Lambda = 0.72$, and a Hubble constant $H_0 = 70\,h_{70}\,\mathrm{km\,s^{-1}\,Mpc^{-1}}$, with $h_{70}=1.0$.

\section{Data}\label{sec:data}

In this Section, we describe the datasets used in our analysis, including the cluster catalogs and the Sloan Digital Sky Survey (SDSS) photometry and spectroscopy data on which our membership selection is based.

\subsection{\textsc{CoMaLit}}\label{sec:data:comalit}

We make use of the \textsc{CoMaLit} compilation \citep{sereno15}, a publicly available meta–catalog of galaxy clusters with weak-lensing mass measurements. The current release (version 3.9) includes more than 1500 mass estimates reported at overdensities of 2500c, 500c, 200c, and virial mass. The catalog brings together different weak-lensing analyses in a homogeneous way, adopting consistent cosmological assumptions and standardized procedures to convert shear profiles into mass estimates. All measurements are rescaled to common overdensity definitions, allowing robust cross-survey comparisons. Throughout this work, we restrict the parent sample to clusters not flagged as part of complex or merging structures.

From this catalog, we select clusters in the mass range $13 < \log(M_{200\mathrm{c}}/{h_{70}}^{-1}M_{\odot}) < 15.5$,
and in the redshift interval $0.01 < z < 0.8$. This broad range allows us to explore potential redshift evolution in the splashback-related quantities discussed in the next sections. We further require clusters to lie within the SDSS DR18\footnote{\url{skyserver.sdss.org/dr18}} footprint, since our membership assignment relies on SDSS spectroscopic and photometric data. SDSS spectroscopy is complete down to $r \simeq 18$ except in regions affected by fiber collisions, while the photometric catalog reaches $r \simeq 22$ \citep{strauss02}.

\begin{figure*}
\gridline{
\fig{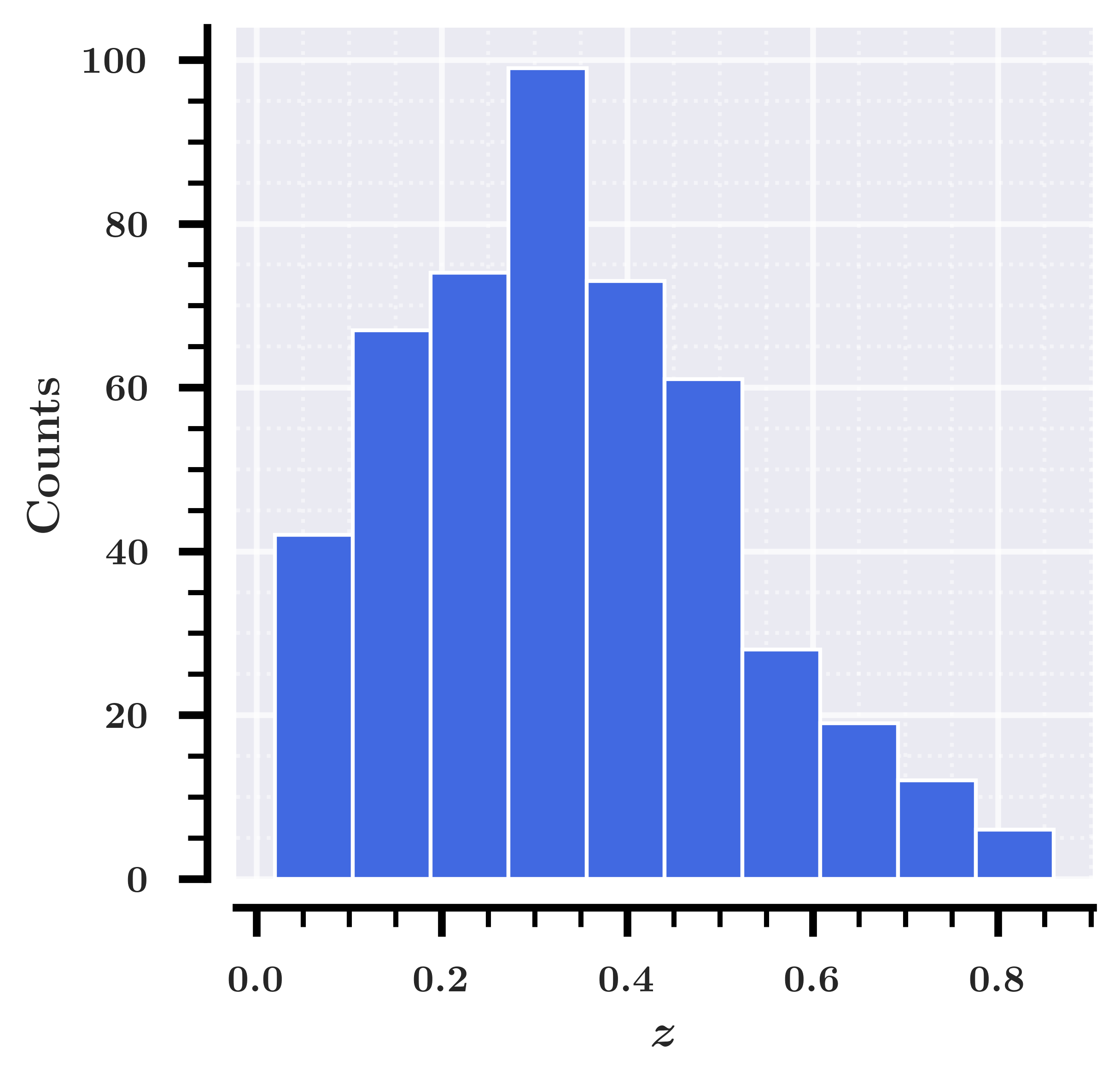}{0.5\textwidth}{(a) Redshift distribution.}
\fig{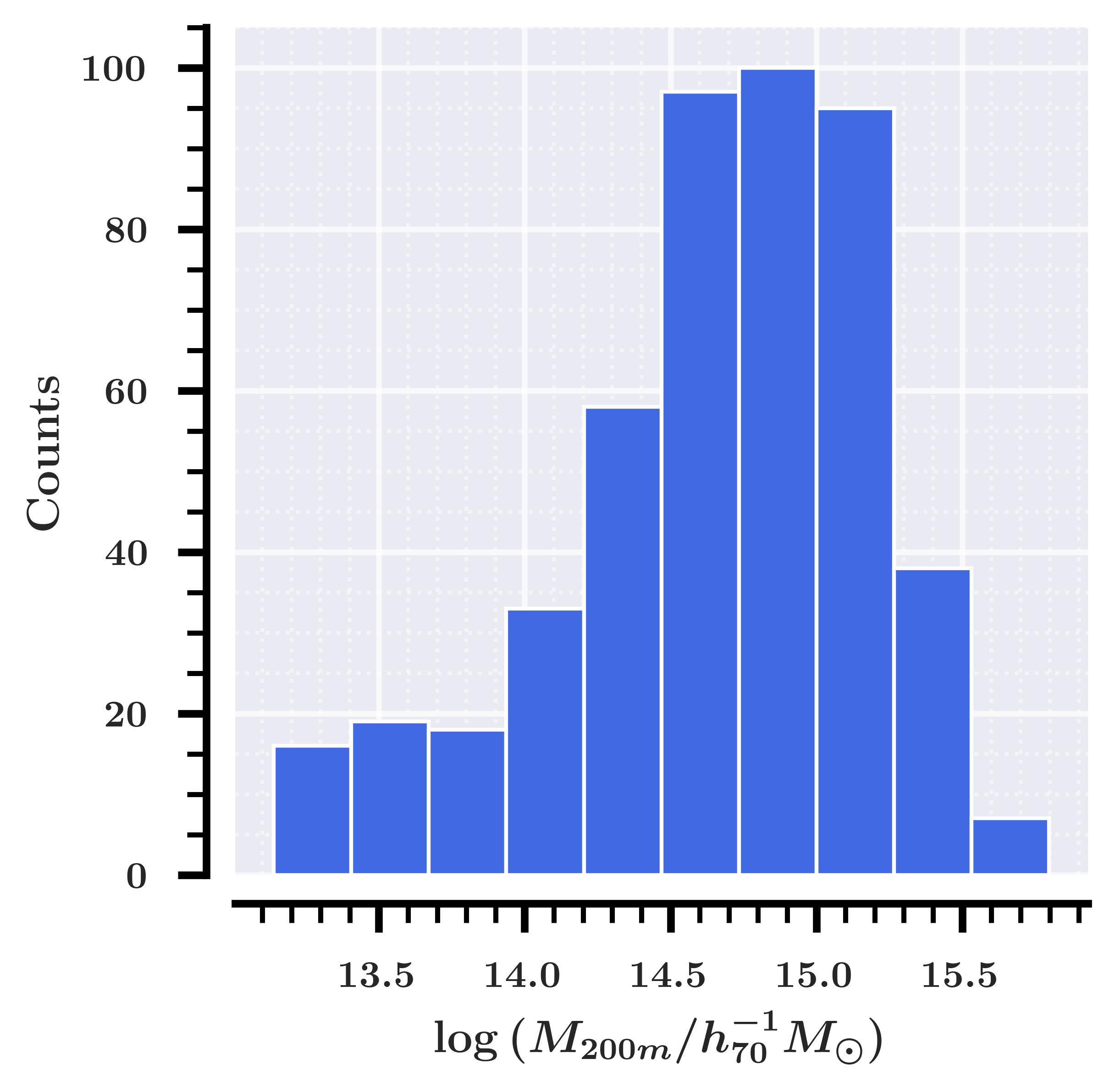}{0.5\textwidth}{(b) Mass distribution.}
}
\caption{Distribution of redshifts (a) and masses (b) for the clusters in \textsc{CoMaLit} that satisfy the selection criteria described in Section~\ref{sec:data:comalit}.}
\label{fig:mass_z_dist}
\end{figure*}

Because splashback studies commonly scale radii by $R_{200\mathrm{m}}$, which leads to more self-similar behavior in the cluster outskirts \citep[e.g.,][]{diemer+14, shi16, umetsu17}, we convert all mass estimates from $200\mathrm{c}$ to $200\mathrm{m}$ using the mass–concentration relation from \citet{diemer19}. After applying all selection criteria described above, and requiring SDSS coverage to extend beyond $3\,R_{200\mathrm{m}}$ in physical Mpc, our final \textsc{CoMaLit} sample contains 499 clusters. Their mass and redshift distributions are shown in Figure \ref{fig:mass_z_dist}.

A smaller subsample of this dataset, about 60 clusters, was previously analyzed in \citet{gabriel-silva+25}. These systems benefit from extensive spectroscopic coverage and are therefore ideal for evaluating the performance of the photometric membership procedure adopted here. This subset is biased toward low redshifts ($0.01 < z < 0.2$) and the high-mass end ($\log(M_{200\mathrm{m}}/{h_{70}}^{-1}M_{\odot}) \approx 14.8$). In the previous work, splashback radii were estimated using spectroscopic members. Here, this spectroscopic sample serves as a validation set for testing how well photometric redshifts alone can recover the membership and the resulting photometric splashback radii ($R_{\mathrm{sp}}^{\mathrm{photo}}$).

For each cluster in both samples, we exclude galaxies flagged as \verb|SATURATED|, \verb|SATUR_CENTER|, \verb|BRIGHT|, or \verb|DEBLENDED_AS_MOVING| in SDSS. We use extinction-corrected model magnitudes and apply k-corrections following the analytical prescriptions from \citet{chilingarian10, chilingarian11}, which introduce typical corrections of $\approx 0.03$ mag in the $r$ band. To maintain consistency across the full redshift range, we only include galaxies brighter than $M_r < -19$ (corresponding to $m_r \approx 22$ at the median redshift of $z=0.3$). This avoids overpopulating the galaxy counts of low-$z$ clusters with faint objects, although previous studies show that this limit does not significantly affect splashback measurements \citep[e.g.,][]{gabriel-silva+25, more+16, murata+20, oneil+22, oshea+24}. Spectroscopy is used only for the subsample analyzed in previous work; in all other cases, cluster membership relies exclusively on photometric redshifts.

\subsection{\textsc{redMaPPer}}\label{sec:data:redmapper}

\begin{figure*}
\centering
\includegraphics[width=1.0\linewidth]{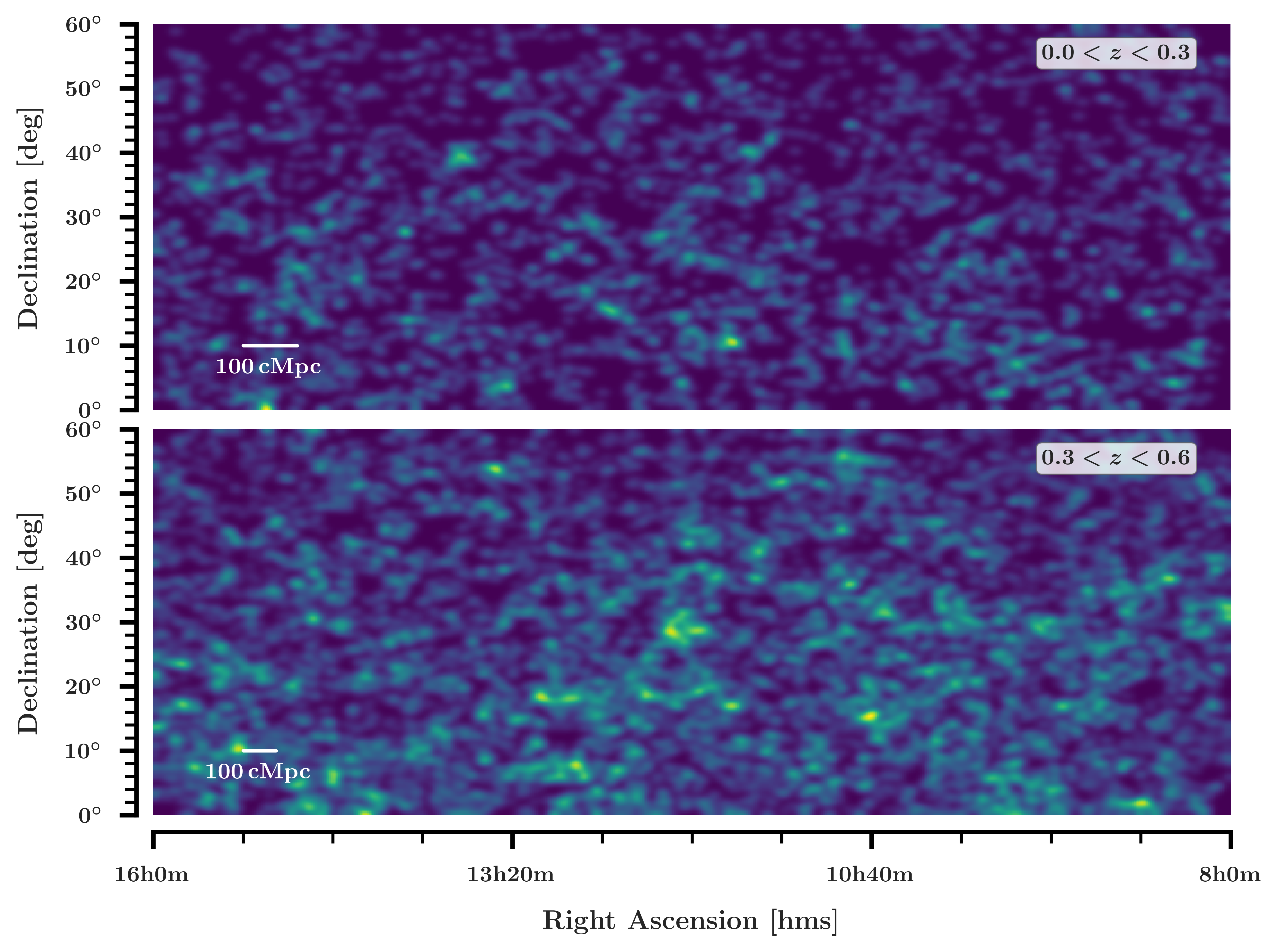}
\caption{\textsc{redMaPPer} clusters sky distribution in the NGC region. Color grade represents surface number density and the horizontal bar indicates a comoving scale of 100 cMpc evaluated at the median redshift of the each sample.}
\label{fig:redmapper_sky}
\end{figure*}

To apply our photometric membership method and splashback estimation to a broader photometric dataset (see Section~\ref{sec:results:mass_func}), we also use the \textsc{redMaPPer} cluster catalog \citep{rykoff+016} constructed from SDSS DR8. \textsc{redMaPPer} identifies galaxy clusters using a red-sequence–based algorithm optimized for large photometric surveys. The catalog spans $0.08 < z < 0.6$ and includes clusters with corrected richness $\lambda/S > 20$, roughly corresponding to
$\log(M_{500\mathrm{c}}/{h_{70}}^{-1}M_{\odot}) \approx 14.0$,
where $S$ is the richness scale factor. To reproduce a well-behaved cluster mass function in a fixed volume, we restrict the sample to the North Galactic Cap (NGC) region ($120^\circ < \mathrm{RA} < 240^\circ$, $0^\circ < \mathrm{DEC} < 60^\circ$). The sky distribution of this subsample is shown in Figure~\ref{fig:redmapper_sky}, where the color scale represents the surface number density, and the horizontal bar indicates a comoving scale of 100 cMpc evaluated at the median redshift of the sample.

Cluster masses are estimated using the richness–mass relation calibrated by \citet{simet+16}. Although the original calibration applies to $z < 0.33$, the relation is commonly extrapolated to higher redshifts and has been widely used in SDSS-based analyses; we follow the same approach here.

Galaxy selection for \textsc{redMaPPer} clusters follows the same criteria adopted for \textsc{CoMaLit}: we use SDSS photometry, exclude objects with problematic flags, and apply the $M_r < -19$ threshold. In this case, all membership assignments rely solely on photometric redshifts.

\section{Methods}\label{sec:methods}

In this Section we describe the probabilistic method developed to assess cluster membership based solely on photometric information, as well as the procedure used to define the optimal probability threshold. We also present the model adopted to fit the cumulative profile, which is crucial for estimating both the splashback radius and the associated splashback mass. Finally, we briefly describe how we construct a galaxy cluster mass function from the splashback masses derived for the \textsc{redMaPPer} catalog.

\subsection{Probabilistic Photometric Membership}\label{sec:methods:prob_memb}

We begin by selecting galaxies with $M_r < -19$ and projected distances within $3\,R_{200\mathrm{m}}$ from the cluster center. We further apply a cut in photometric redshift, requiring galaxies to fall within $\pm 3\sigma_0$ of the cluster redshift, where $\sigma_0 \simeq 0.0205$ is the SDSS normalized median absolute deviation \citep{strauss02}. 

From this initial set, we define a core radius
\[
R_\mathrm{core} = \frac{R_{200\mathrm{m}}}{2},
\]
which is used to model the cluster red sequence (RS). Galaxies within $R_\mathrm{core}$ are fit with a linear RS model, and we exclude objects lying more than $3\sigma_{\mathrm{RS}}$ above the best-fit relation. The value of $\sigma_{\mathrm{RS}}$ is taken from the fit, but we impose bounds of $0.05$--$0.15$\,mag to avoid extreme values. Blue-cloud galaxies are retained.

We also estimate the magnitude of the third brightest galaxy within $R_\mathrm{core}$ ($M_3$) and exclude galaxies outside $R_\mathrm{core}$ that are brighter than $M_3 + \Delta_m$, adopting $\Delta_m \simeq 0.5$\,mag. Although bright galaxies may exist in cluster outskirts, it is well established that the brightest population is strongly concentrated toward the center in relaxed systems \citep[e.g.,][]{lin+04, hoshino+15}. Therefore, this cut efficiently removes obvious contaminants. The resulting sample constitutes the clean dataset used for the probabilistic membership inference.

To assign photometric memberships, we adopt a Bayesian model adapted from L\"osch et al. (in prep.), in which the membership probability is the product of two independent components: a redshift term and a radial term.

\subsubsection{Redshift Probability}

The redshift component of the membership probability is defined as
\begin{equation}
    P_z = \int_{z_c - 1.5\,\sigma_z(m)}^{z_c + 1.5\,\sigma_z(m)} PDF(z)\,dz,
\end{equation}
where $z_c$ is the cluster redshift, and $PDF(z)$ is the photometric-redshift probability density function. For SDSS, we assume a Gaussian form based on the reported photo-$z$ and its uncertainty. The quantity $\sigma_z(m)$ is the magnitude-dependent photo-$z$ scatter.

\begin{figure}
\centering
\includegraphics[width=1.0\linewidth]{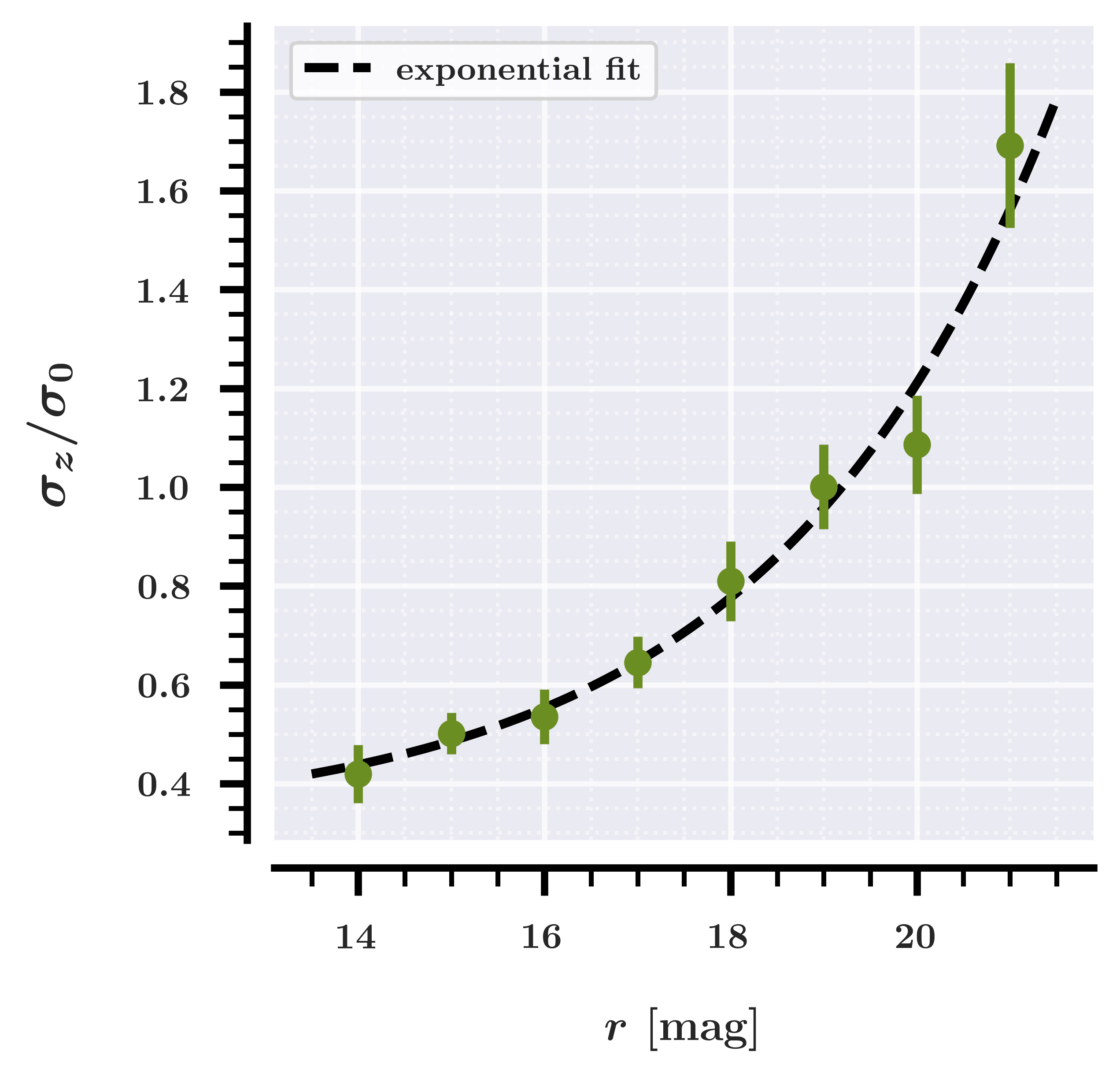}
\caption{Magnitude dependence of the SDSS photo-$z$ dispersion in the $r$ band.}
\label{fig:sdss_sigmaz}
\end{figure}

Although SDSS provides a global $\sigma_0 \simeq 0.02$, it is well known that photo-$z$ performance degrades with magnitude \citep{lima+22}. To model this dependence, we compiled $\sim 150{,}000$ SDSS galaxies with both spec-$z$ and photo-$z$, and fit the magnitude dependence of the scatter using an exponential function. The resulting relation is shown in Figure~\ref{fig:sdss_sigmaz}, where error bars reflect a bootstrap estimation. The dispersion is normalized by the SDSS absolute error $\sigma_0$. This procedure links the redshift probability to both the galaxy magnitude and the intrinsic photo-$z$ uncertainty.

The factor of $1.5$ multiplying $\sigma_z(m)$ is adopted following L\"osch et al. (in prep.), who found it maximizes completeness and purity in mock simulations. This is also consistent with the fact that clusters detected in photometric surveys typically display redshift dispersions of order $\sigma_z/2$ \citep{castigani+16}.

\subsubsection{Radial Probability}

For the radial component, we estimate the galaxy surface-number-density profile from the clean sample described above. At this stage, contamination is still significant, so to suppress interlopers we weight each galaxy by its redshift probability $P_z$ when computing the density profile. We then model the resulting profile using a Sérsic component plus a constant background:
\begin{equation}
    \Sigma_{\mathrm{model}}(R) = \Sigma_s(R) + \Sigma_{\mathrm{bkg}},
\end{equation}
where
\begin{equation}
    \Sigma_s(R) = \Sigma_e \exp\left[-b_n\left( \left(\frac{R}{R_e}\right)^{1/n} - 1\right)\right],
\end{equation}
with $R_e$ the effective radius, $\Sigma_e$ the density at $R_e$, and $n$ the Sérsic index.

This simple model may seem insufficient given that splashback studies require a sharp truncation in the outskirts \citep[e.g.,][]{diemer+14, baxter+17}. However, fitting a fully truncated profile at this stage is not feasible because residual contamination strongly biases the inferred parameters. Even in spectroscopy-based samples, increasing the redshift window can bias $R_{\mathrm{sp}}$ downward by $5$--$10\%$ \citep[e.g.,][]{gabriel-silva+25}, and we verify that photo-$z$ uncertainties can increase this to $20$--$50\%$. The Sérsic$+$background model is therefore used only as an intermediate step to remove contaminants. A more sophisticated truncated profile is fit later when estimating the splashback radius.

The radial membership probability is then defined as:
\begin{equation}
    P_r = \frac{\Sigma_s}{\Sigma_s + \Sigma_{\mathrm{bkg}}}.
\end{equation}

\subsubsection{Probability Cut}

The total membership probability is given by
\begin{equation}
    P_f = P_z\,P_r.
\end{equation}

For numerical stability, we combine these probabilities using their odds representation:
\begin{equation}
    P_f = \frac{\mathrm{odds}}{1 + \mathrm{odds}},
\end{equation}
where
\begin{equation}
    \mathrm{odds} = 
    \exp\!\left[
    \ln\!\left(\frac{P_z}{1-P_z}\right) +
    \ln\!\left(\frac{P_r}{1-P_r}\right)
    \right].
\end{equation}

Instead of adopting a fixed hard cut (e.g. $P_f > 0.5$), we implement an adaptive threshold that adjusts to cluster richness. Rich clusters require a stricter cut, while poor clusters benefit from a softer one. To define the optimal threshold $P_{\mathrm{cut}}$, we maximize the detection significance within the core region ($N_\mathrm{core}$) relative to the outer region ($N_{\mathrm{out}}$) between $R_{200\mathrm{m}}$ and $3R_{200\mathrm{m}}$.

Following \citet{lima83}, the significance is
\begin{equation}\label{eq:SN}
\begin{aligned}
\mathrm{SN} = \sqrt{2}
\left[
N_{\mathrm{core}}
\ln\!\left(
\frac{1+\alpha}{\alpha}
\frac{N_{\mathrm{core}}}{N_{\mathrm{core}}+N_{\mathrm{out}}}
\right)
\right.
\\[6pt]
\left.
\quad +
N_{\mathrm{out}}
\ln\!\left(
(1+\alpha)
\frac{N_{\mathrm{out}}}{N_{\mathrm{core}}+N_{\mathrm{out}}}
\right)
\right]^{1/2}.
\end{aligned}
\end{equation}
where the effective number of galaxies is estimated as
\begin{equation}
N = 
\frac{\left(\sum P_f\right)^{2}}
{\sum P_f^{2}},
\end{equation}
and the geometric exposure ratio is
\begin{equation}
\alpha = 
\frac{\pi R_\mathrm{core}^{2}}
{\pi \left[(3R_{200\mathrm{m}})^{2} - R_{200\mathrm{m}}^{2}\right]}.
\end{equation}

We evaluate $\mathrm{SN}$ on a grid of $P_{\mathrm{cut}} \in [0, 1]$ in steps of $0.01$. The optimal threshold is defined as
\begin{equation}
P_\mathrm{cut}^{\mathrm{best}} = 
\underset{P_\mathrm{cut}}{\arg\max}\;
\mathrm{SN}\!\left( N_{\mathrm{core}}(P_\mathrm{cut}), 
N_{\mathrm{out}}(P_\mathrm{cut}), 
\alpha \right).
\end{equation}

To assess the performance of the photometric probabilistic membership, we use our subsample of 60 \textsc{CoMaLit} clusters with extensive spectroscopic coverage. Spectroscopic members are identified using the sigma-Gapper method called CausticSNUpy from \citet{kang+24}, which infers an empirical caustic based on the galaxy redshifts and projected distances from the cluster center. The code first
finds candidate members using a hierarchical clustering
algorithm. The code then calculates the galaxy number
density in the redshift space via adaptive kernel density
estimation. For this method, we fix the cluster centers using their known RA, Dec, and redshift, rather than allowing the algorithm to infer them.

\subsection{Splashback Feature}\label{sec:methods:sp_feature}

Following the same approach as in our previous work \citep{gabriel-silva+25}, we model the cumulative number profile of galaxy clusters to recover their density profiles and identify the splashback feature. Although splashback radii are typically estimated from the surface-number-density profile, which traces the dark matter density reasonably well \citep[e.g.,][]{lebeau+24}, the cumulative distribution is considerably more stable, especially in systems with low galaxy counts such as poor or high-redshift clusters. In this sense, estimating the splashback radius for individual clusters becomes significantly more robust.

\subsubsection{Cumulative Model}

Assuming spherical symmetry, the cumulative profile is obtained from the surface-number-density profile through
\begin{equation}\label{eq:N(R)}
    N(<R) = 2\pi \int_0^R \Sigma(R)\, R\, dR,
\end{equation}
where the surface density $\Sigma(R)$ follows the form proposed by \citet{diemer+14}, consisting of a one-halo term, a two-halo term, and a smooth truncation function. A more recent formulation \citep{diemer22} introduces additional flexibility and has been preferred in recent works; however, it involves a larger number of free parameters, which in cumulative-profile analyses leads to increased degeneracies. We therefore adopt the so-called trunc-NFW model, which was previously shown to work well for individual cluster estimates \citep{gabriel-silva+25}:

\begin{equation}
    \Sigma(R) = \Sigma_{\mathrm{NFW}}\, f_t + \Sigma_{\mathrm{power-law}}.
\end{equation}

The one-halo component is described by the analytically projected Navarro--Frenk--White (NFW) profile \citep{navarro+96, merritt+05} in the formulation of \citet{lokas+01}:
\begin{align}
    \Sigma_{\mathrm{NFW}}(R) 
    &= \frac{2 r_s \rho_s}{x^2 - 1}
       \left( 1 - \frac{\mathcal{C}(x)}{\sqrt{|x^2 - 1|}} \right), \\
    \mathcal{C}(x) &= 
    \begin{cases}
        \cosh^{-1}(1/x), & x < 1, \\
        \cos^{-1}(1/x), & x > 1,
    \end{cases} \\
    x &= \frac{R}{r_s}.
\end{align}

The truncation function $f_t$ follows the smooth exponential form of \citet{diemer22}:
\begin{equation}
    f_t = \exp\left[ -\left(\frac{R}{R_t}\right)^{\tau} \right].
\end{equation}

The two-halo contribution is modeled using the projected correlation function derived by \citet{davis+83}, expressed as a power law:
\begin{equation}
    \Sigma_{\mathrm{power-law}} = 
    \rho_m \left[
    \left( \frac{R}{r_{\mathrm{out}}} \right)^{1 - \gamma}
    B\!\left( \frac{\gamma - 1}{2}, \frac{1}{2} \right)
    \right],
\end{equation}
where $B(X,Y)$ is the Beta function. We fix $r_{\mathrm{out}} = 1.5$ Mpc due to its degeneracy with~$\rho_m$.

The full trunc-NFW model contains six free parameters:
\[
(\rho_s, r_s, R_t, \tau, \rho_m, \gamma).
\]
We fit this model using an MCMC sampler with a Poisson likelihood, adopting the priors listed in Table~\ref{tab:priors}. To avoid numerical issues commonly associated with Abel inversions, we use fully analytical projected expressions. The splashback radius is then definied as
\begin{equation}
R_\mathrm{sp} = 
\underset{R}{\arg\min}\;
\frac{d\log{\Sigma(R)}}{d\log{R}}.
\end{equation}

As discussed in \citet{walker+25}, the truncation radius, $R_t$, provides a direct measure of the edge of the orbiting material, whereas the splashback radius, $R_{\rm sp}$, defined as the radius of steepest slope, reflects the transition between orbiting and infalling components. While the two radii coincide in ideal spherical collapse, realistic anisotropic mass accretion smooths the truncation of the orbiting profile, causing $R_{\rm sp}$ to lie systematically outside $R_t$.

\begin{deluxetable}{ccc}
\tabletypesize{\scriptsize}
\tablewidth{0pt}
\tablecaption{Priors used in the trunc-NFW model for fitting cumulative number profiles.\label{tab:priors}}
\tablehead{
\colhead{Parameter} & \colhead{Value} & \colhead{Type}
}
\startdata
$\rho_s$ & 0--500 & Uniform \\
$r_s$ & 0.1--1.5 Mpc & Uniform \\
$R_t$ & 0.1--8.0 Mpc & Uniform \\
$\tau$ & $4 \pm 0.8$ & Gaussian \\
$\rho_m$ & 0--100 & Uniform \\
$r_{\mathrm{out}}$ & 1.5 Mpc & Fixed \\
$\gamma$ & $1.7 \pm 0.4$ & Gaussian \\
\enddata
\end{deluxetable}

\subsubsection{Splashback Mass}

To estimate the splashback mass, we begin with the tabulated \textsc{CoMaLit} $M_{200\mathrm{c}}$ values and extend the mass profile using the NFW expression from \citet{lokas+01} combined with the same truncation function:
\begin{align}\label{eq:M_r}
    M(<r) &= 
    4\pi \int_0^r 
    \frac{r^2 M_{200\mathrm{c}}}{4\pi r_s^3 g_c}
    \frac{f_t}{x(1+x)^2}\, dr, \\
    g_c &= \log(1 + c_{200\mathrm{c}}) - \frac{c_{200\mathrm{c}}}{1 + c_{200\mathrm{c}}}, \\
    c_{200\mathrm{c}} &= \frac{R_{200\mathrm{c}}}{r_s}.
\end{align}

Using the fitted parameters $(r_s, R_t, \tau)$ and the cataloged $M_{200\mathrm{c}}$, we compute the mass enclosed within any radius. In particular, the splashback mass is defined as
\begin{equation}\label{eq:M_sp}
    M_{\mathrm{sp}} \equiv M(<R_{\mathrm{sp}}).
\end{equation}

The truncation function does not significantly modify mass estimates at radii much smaller than $R_{\mathrm{sp}}$, consistent with \citet{giocoli+24, gabriel-silva+25}. Since $R_t$ is typically comparable to $R_{\mathrm{sp}}$ and both lie well beyond $R_{200\mathrm{c}}$, the mass within the inner region remains essentially unaffected. We also do not include the two-halo contribution in this definition, as its impact within $R_{\mathrm{sp}}$ is small ($\sim 1$--$5\%$) and negligible compared to observational uncertainties.

It is important to note that, in this procedure, we combine projected (2D) determinations of the splashback feature with a three-dimensional mass profile, which may introduce a small projection-related bias. However, previous tests indicate that the intrinsic uncertainties in splashback mass estimates dominate over this effect \citep{gabriel-silva+25}. Therefore, this method provides a reasonable and practical estimate of $M_{\mathrm{sp}}$ given current observational limitations.

For the splashback mass uncertainty, we consider errors in both $M_{200\mathrm{c}}$ and $R_{\mathrm{sp}}$. We assume these contributions are independent and propagate them separately: the uncertainty from $M_{200\mathrm{c}}$ through Equation~\ref{eq:M_sp}, denoted $\sigma_{M_{\mathrm{sp}}, M_{200\mathrm{c}}}$, and the uncertainty from the posterior distribution of $R_{\mathrm{sp}}$, denoted $\sigma_{M_{\mathrm{sp}}, R_{\mathrm{sp}}}$. The total error is then
\begin{equation}
    \sigma_{M_{\mathrm{sp}}}
    = \sqrt{
    \sigma_{M_{\mathrm{sp}}, M_{200\mathrm{c}}}^{2}
    +
    \sigma_{M_{\mathrm{sp}}, R_{\mathrm{sp}}}^{2}}.
\end{equation}

Our results show that the total uncertainty is dominated by the propagation from $M_{200\mathrm{c}}$. Typical contributions are $\sigma_{M_{\mathrm{sp}},\,M_{200\mathrm{c}}} \simeq 0.2$ dex, while the contribution associated with the uncertainty in the splashback radius is significantly smaller, $\sigma_{M_{\mathrm{sp}},\,R_{\mathrm{sp}}} \simeq 0.05$ dex.

\subsection{Cluster Mass Function}\label{sec:methods:mass_func}

Our primary motivation for estimating the splashback mass is to refine and apply the $R_{\mathrm{sp}}$--$M_{\mathrm{sp}}$ relation introduced in our previous work \citep{gabriel-silva+25}. Once this relation is calibrated, it provides a direct mapping between the splashback radius and the corresponding cluster mass, enabling straightforward mass estimates from measurements of $R_{\mathrm{sp}}$. A natural and robust way to assess the consistency of these estimates is through the construction of a galaxy cluster mass function defined in terms of $M_{\mathrm{sp}}$. The abundance of dark matter halos as a function of mass is one of the most fundamental statistical quantities in cosmology, establishing a direct connection between theoretical models and observational data, particularly when cluster counts are used to constrain cosmological parameters. Classical descriptions of this distribution are provided by analytical models such as \citet{press+74}, in which peaks in a Gaussian random field collapse once their overdensity exceeds a critical threshold.

To estimate the observational splashback mass function, we select all \textsc{redMaPPer} clusters within the SDSS NGC region and compute their photometric membership probabilities and splashback radii following the methodology described in Sections~\ref{sec:methods:prob_memb} and~\ref{sec:methods:sp_feature}. For each cluster, the measured splashback radius $R_{\mathrm{sp}}$ is converted into a splashback mass $M_{\mathrm{sp}}$ using the best-fit scaling relation obtained in Section~\ref{sec:results:sp_feature}. The sample is divided into two redshift bins, $0.08 < z < 0.3$ and $0.3 < z < 0.6$, and for each bin we compute the maximum comoving volume corresponding to the surveyed sky area and redshift range. We then bin the splashback masses into ten logarithmic intervals for clusters with $M_{\mathrm{sp}} > 10^{14}\,h_{70}^{-1}M_{\odot}$, and obtain the mass function by dividing the number of clusters in each bin by the corresponding comoving volume.

This procedure does not explicitly account for the \textsc{redMaPPer} selection function and therefore implicitly assumes approximate completeness above the adopted mass threshold. Potential incompleteness effects, particularly at lower masses and higher redshifts, are not corrected for here and are discussed in more detail in Section~\ref{sec:results:mass_func}.

This procedure yields an observational galaxy cluster mass function defined in terms of the splashback mass. For comparison, we adopt the analytical formulations presented by \citet{diemer20}, who calibrated halo mass functions expressed in terms of $M_{\mathrm{sp}}$ using cosmological simulations. In this framework, the splashback feature is not associated with a single, sharply defined radius. Instead, recently accreted dark matter particles reach their first apocenter over a broad range of radii, resulting in a distribution of splashback radii rather than a unique value \citep[e.g.,][]{diemer20, diemer21}. This distribution can be characterized through splashback quantiles, defined as percentiles of the radial distribution of first apocenters, which give rise to different effective halo boundaries and, consequently, to a family of splashback mass functions. In observational data, the splashback quantile cannot be directly determined; instead, our methodology is expected to recover an effective average behavior, likely close to the median of the distribution, corresponding approximately to the $50\%$ splashback quantile.

\section{Results}\label{sec:results}

In this Section, we present and discuss the main outcomes of our analysis, ranging from the performance of the probabilistic membership estimator to its impact on the recovery of the splashback feature and final mass function results.

\subsection{Probabilistic Photometric Membership}\label{sec:results:prob_memb}

As described in Section~\ref{sec:methods:prob_memb}, our membership estimator combines independent radial and redshift probability components. To quantify its performance, we make use of the \textsc{CoMaLit} spectroscopic subsample, which provides high-quality spec-$z$ coverage for $\sim60$ clusters. Real members are defined using the caustic-based classification from \citet{kang+24}. Although caustic methods rely on projected phase-space and may not fully recover the complex three-dimensional dynamical structure of clusters, this approach is adequate for validation since our primary goal is not strict member identification, but rather recovering a reliable cumulative density profile for splashback measurements.

This validation step is supported by mock tests from L\"osch et al. (in preparation), who applied the same probabilistic algorithm to synthetic galaxy catalogs from \citet{araya-araya21} based on the SAM models of \citet{henriques15}, generated over a projected area of 324 deg$^2$ using the \textsc{Millennium Run} simulation \citep{springel05} rescaled to \textit{Planck}-1 cosmology \citep{planck14}. In the simulated environment, the method reached completeness and purity values of $\sim70\%$. Here, we test whether similar performance holds in real observations.

\subsubsection{Completeness and Purity}

To evaluate membership recovery, we select all galaxies within the spatial footprint of the 60 clusters that possess both photometric and spectroscopic redshifts. A galaxy is considered a real member if it lies within the spectroscopic caustic. Interlopers are those outside the caustic limits.

We then apply our probabilistic photo-$z$ method to the same galaxy set, classifying a galaxy as a probable member if its final probability exceeds a threshold $P_{\mathrm{cut}}$. We define completeness following \citet{fawcett06}:
\begin{equation}
    C(P_{\mathrm{cut}}) = \frac{TP}{TP + FN},
\end{equation}
where $TP$ are real members selected by the probabilistic method and $FN$ are real members missed by it.

Similarly, purity is defined as:
\begin{equation}
    P(P_{\mathrm{cut}}) = \frac{TP}{TP + FP},
\end{equation}
where $FP$ are interlopers incorrectly selected as probable members.

\begin{figure}
\centering
\includegraphics[width=1.0\linewidth]{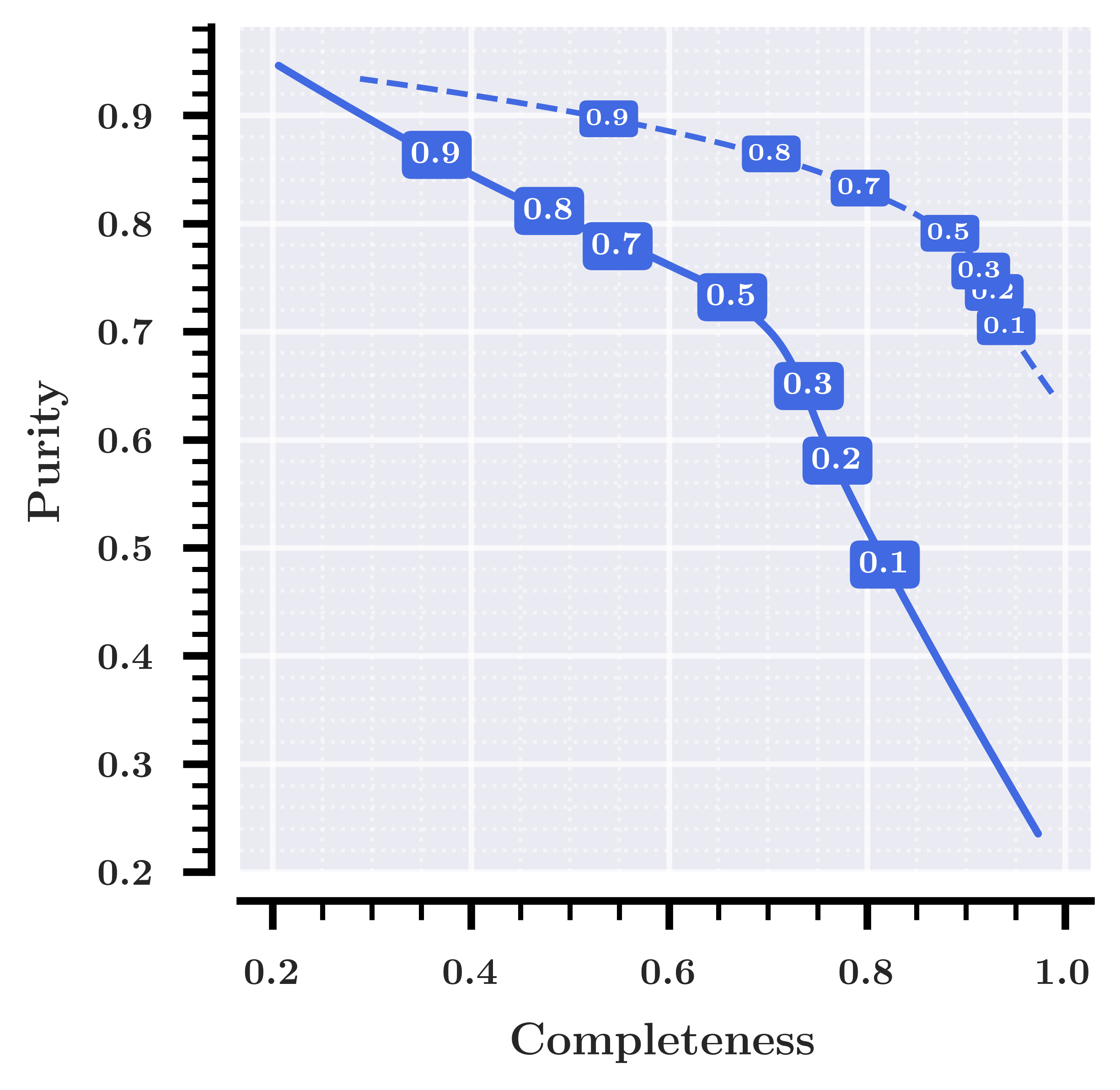}
\caption{Completeness versus purity as a function of the probability threshold. Solid lines show results up to $3R_{200\mathrm{m}}$; dashed lines show results restricted to $R_{200\mathrm{m}}$.}
\label{fig:comp_purity}
\end{figure}

Figure~\ref{fig:comp_purity} shows completeness versus purity as a function of $P_{\mathrm{cut}}$. As expected, higher completeness leads to lower purity, but the trade-off remains favorable. For a typical cut of $P_{\mathrm{cut}}>0.5$, the method reaches $\sim70\%$ completeness and purity up to $3\,R_{200\mathrm{m}}$. Restricting the analysis to $R_{200\mathrm{m}}$ (dashed curve) improves both metrics, reaching $\sim85\%$ purity and $\sim90\%$ completeness, in close agreement with studies adopting similar approaches, such as \citet{castigani+16}, which also include radial and redshift components.

\subsubsection{Adaptive Probability Cut}

\begin{figure}
\centering
\includegraphics[width=1.0\linewidth]{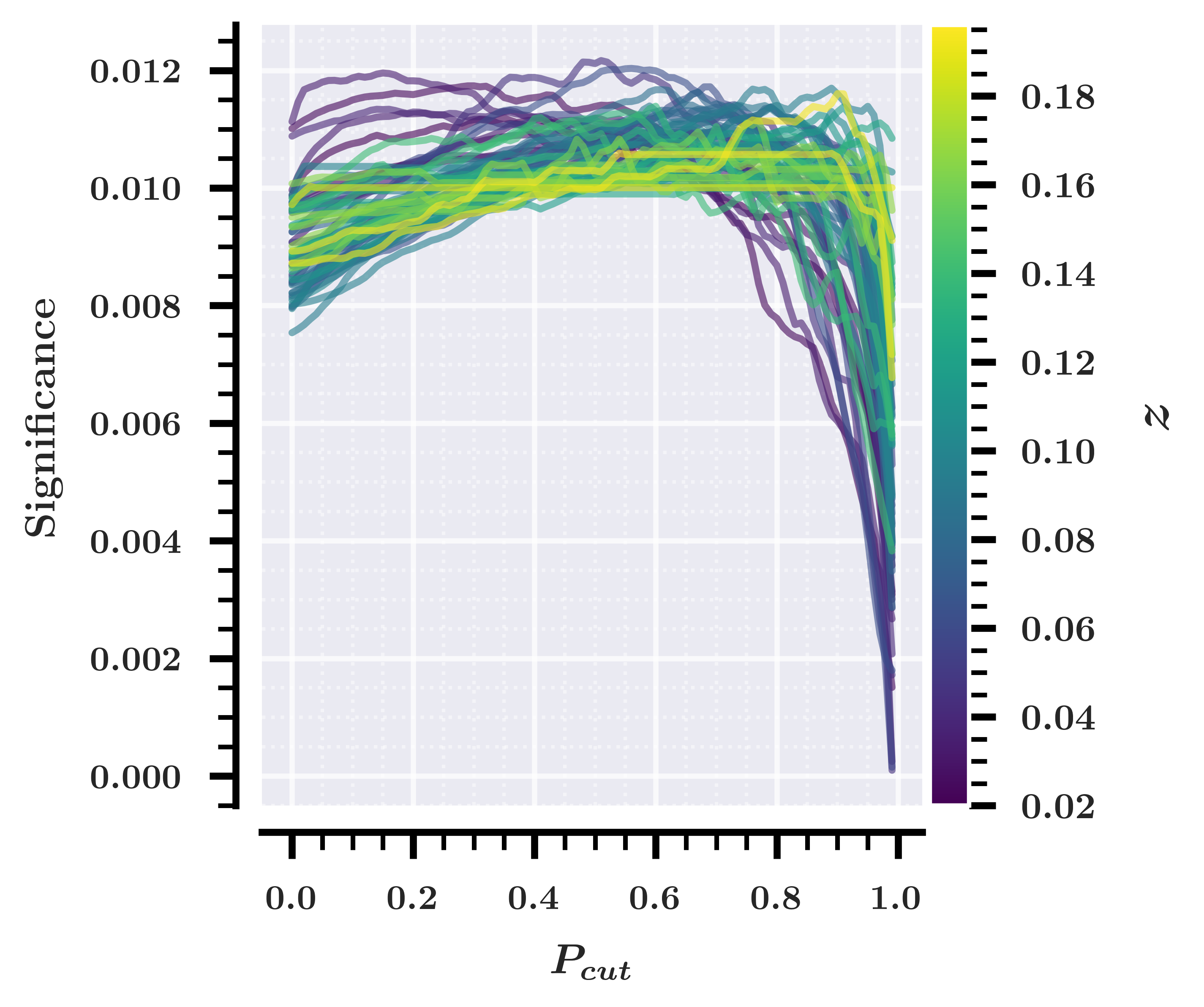}
\caption{Variation of the normalized significance with probability threshold for the \textsc{CoMaLit} subsample. The color gradient indicates redshift.}
\label{fig:SN_Pcut}
\end{figure}

Instead of applying a universal threshold, we determine an individual $P_{\mathrm{cut}}$ for each cluster by maximizing the signal-to-noise ratio defined following \citet{lima83} in Equation~\ref{eq:SN}. Figure~\ref{fig:SN_Pcut} illustrates the relation between significance and $P_{\mathrm{cut}}$ for the 60 clusters, normalized for visualization. A clear trend emerges: most clusters reach maximum significance at $P_{\mathrm{cut}}\sim0.6$. However, when considering the full \textsc{CoMaLit} sample, the median optimal threshold decreases toward $P_{\mathrm{cut}}\sim0.4$, mainly due to increasing redshift leading to lower galaxy counts as consequence of the Malmquist bias in flux-limited samples \citep[e.g.,][]{malmquist22, malmquist25, teerikorpi97}.

\subsubsection{Impact on Splashback Measurements}

\begin{figure}
\centering
\includegraphics[width=1.0\linewidth]{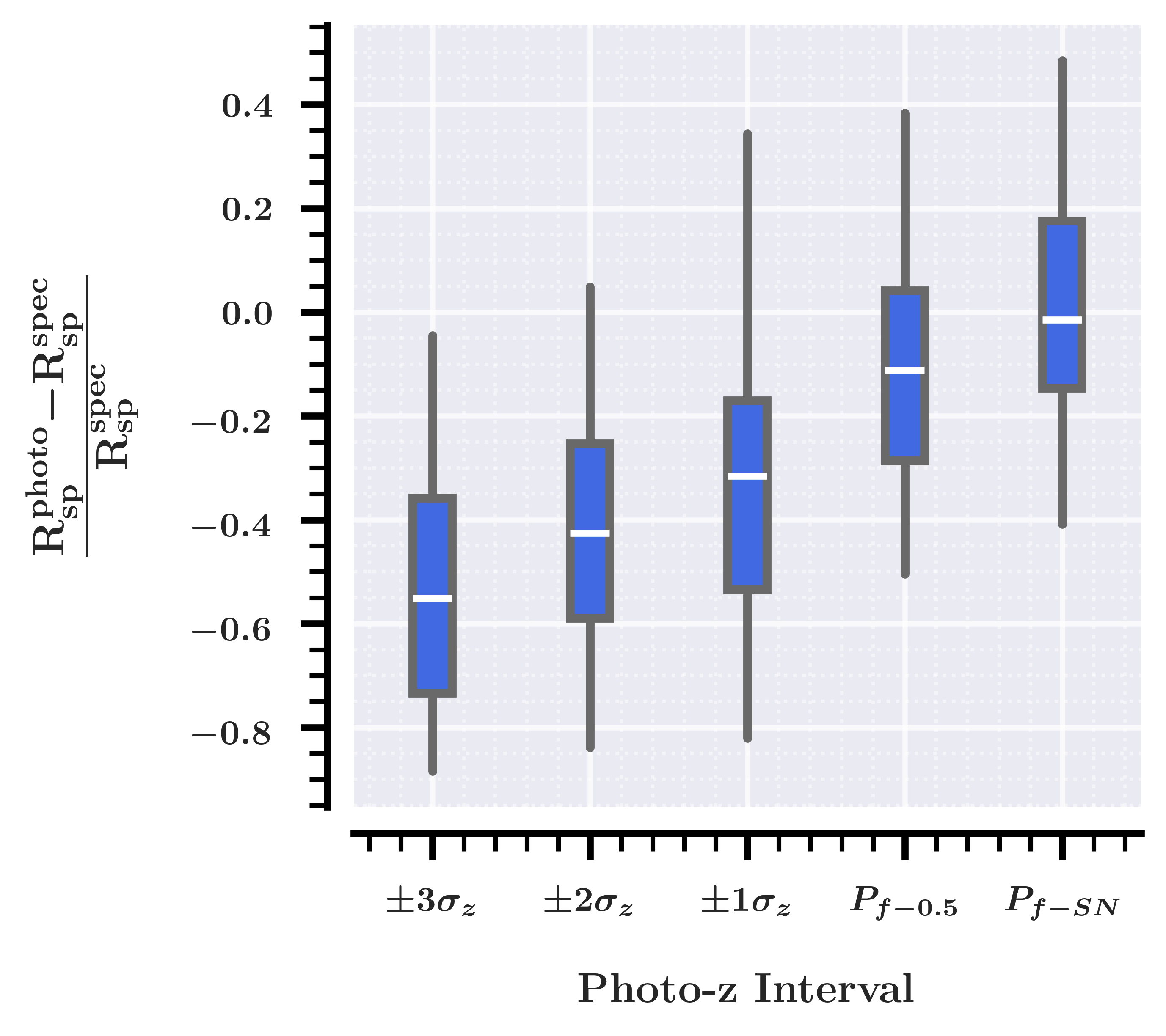}
\caption{Percent error in splashback radius estimates obtained using different photometric membership definitions, relative to the spectroscopic benchmark from \citet{gabriel-silva+25}. The photo-$z$ aperture method selects galaxies within a redshift window $|z - z_c| < i\,\sigma_z$, with $i = 1, 2,$ and $3$ corresponding to increasingly broader selection intervals. The hard-cut probabilistic method ($P_{f-0.5}$) includes galaxies with membership probability $P_f > 0.5$, while the adaptive method ($P_{f\text{-}SN}$) applies a signal-to-noise–optimized threshold.}
\label{fig:spec_photo_sp}
\end{figure}

Figure~\ref{fig:spec_photo_sp} compares the percentage error in $R_{\mathrm{sp}}^{\mathrm{photo}}$ obtained through different photometric membership definitions relative to the spectroscopic benchmark from \citet{gabriel-silva+25}, denoted as $R_{\mathrm{sp}}^{\mathrm{spec}}$. The photo-$z$ aperture technique ($|z - z_c| < i\,\sigma_z$, where $i = 1, 2,$ and $3$) shows increasing bias with larger redshift intervals, consistently underestimating the splashback radius.

In contrast, the probabilistic method significantly reduces systematic bias, particularly when using the adaptive threshold. The hard-cut method ($P_{f-0.5}$) already improves upon the redshift-window approach, while the adaptive method ($P_{f\text{-}SN}$) nearly eliminates the mean bias. However, the scatter remains similar across methods, with a typical dispersion of $\sim20\%$, indicating that most of the measurement uncertainty is intrinsic rather than driven by the membership technique itself. Hereafter, all splashback measurements are photometrically derived, and we adopt $R_{\mathrm{sp}} \equiv R_{\mathrm{sp}}^{\mathrm{photo}}$ for simplicity.

\subsection{Splashback Feature}\label{sec:results:sp_feature}

One of the main results of our previous work \citep{gabriel-silva+25} was the scaling relation between splashback mass and radius, $M_{\mathrm{sp}}$--$R_{\mathrm{sp}}$. However, due to the limited sample size and narrow redshift range available at that time, it was not possible to robustly constrain all parameters of the relation, particularly the redshift dependence. Here, we revisit this analysis using the full \textsc{CoMaLit} sample with SDSS photometry, substantially increasing both the statistical power and redshift leverage.

For each cluster, we apply the probabilistic membership methodology described in Section~\ref{sec:methods:prob_memb}, determine the optimal adaptive probability cut, and select the most probable member galaxies. From these members, we construct the cumulative number profile and model it following Equation~\ref{eq:N(R)}. We sample the model parameters using MCMC and derive the splashback radius for each system. The corresponding splashback mass is then estimated using Equation~\ref{eq:M_sp}, combining the measured $R_{\mathrm{sp}}$, the fitted profile parameters, and the catalog $M_{200\mathrm{c}}$.

\begin{figure*}
\gridline{
\fig{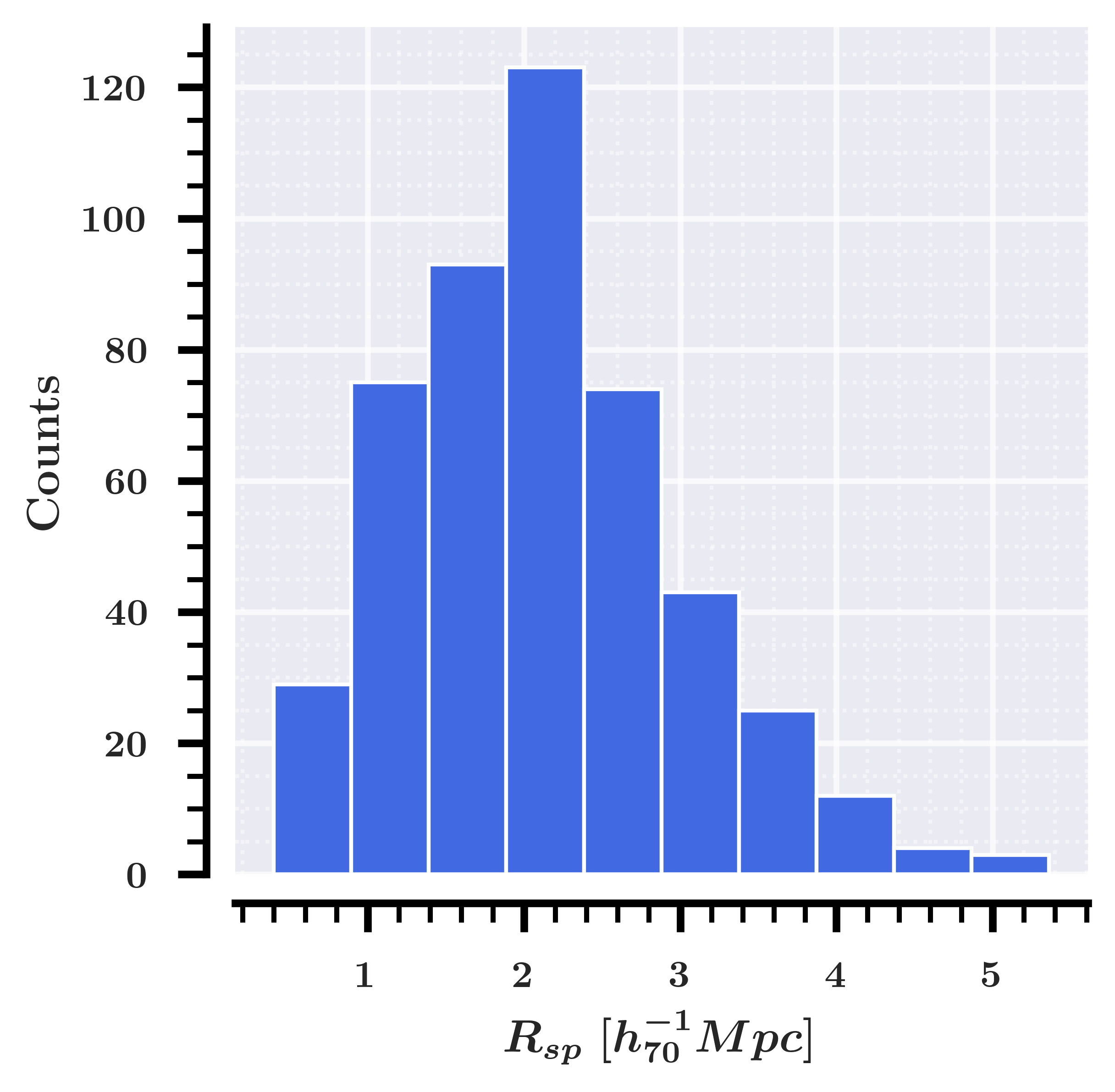}{0.5\textwidth}{(a) Distribution of splashback radii.}
\fig{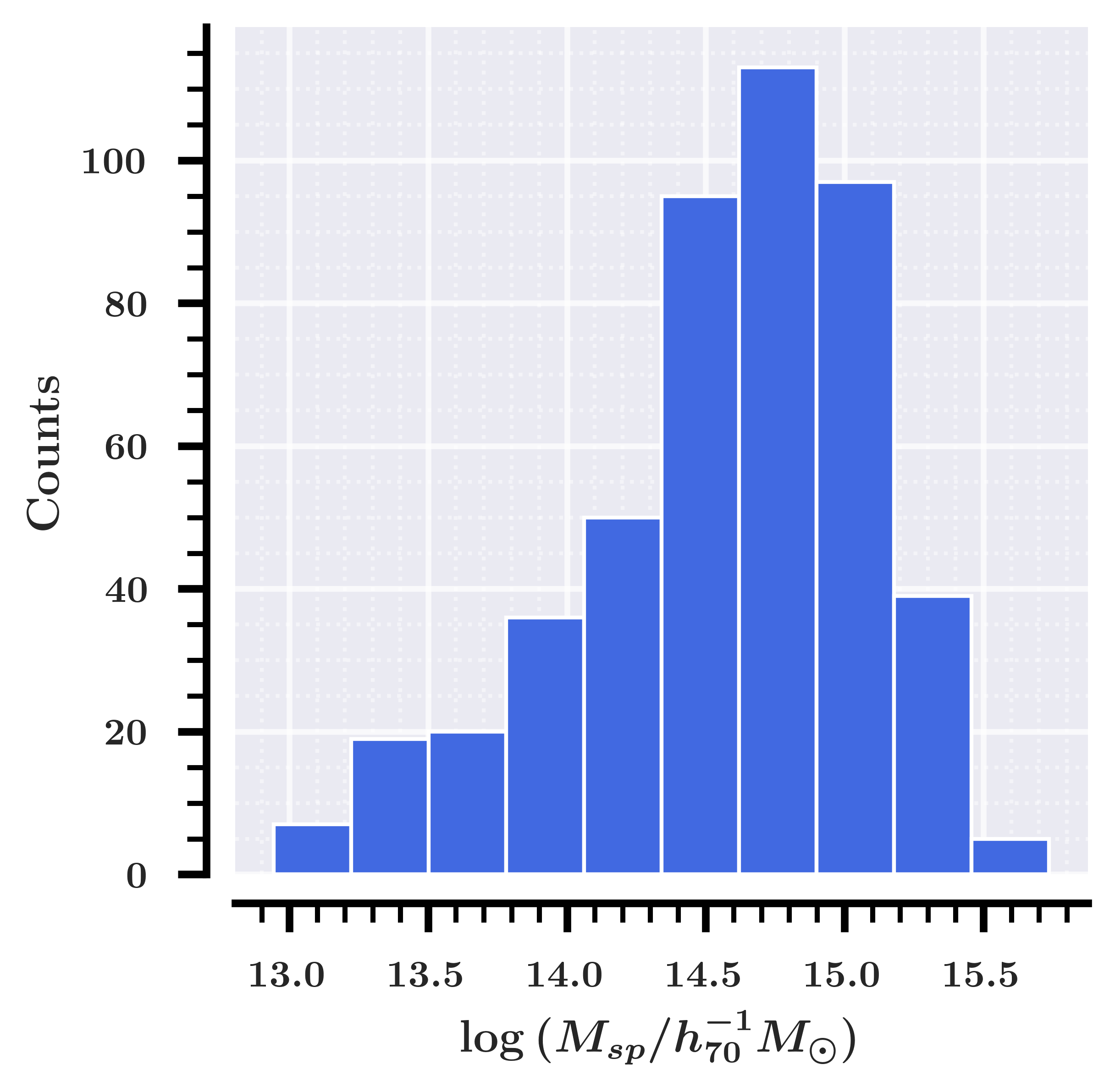}{0.5\textwidth}{(b) Distribution of splashback masses.}
}
\gridline{
\fig{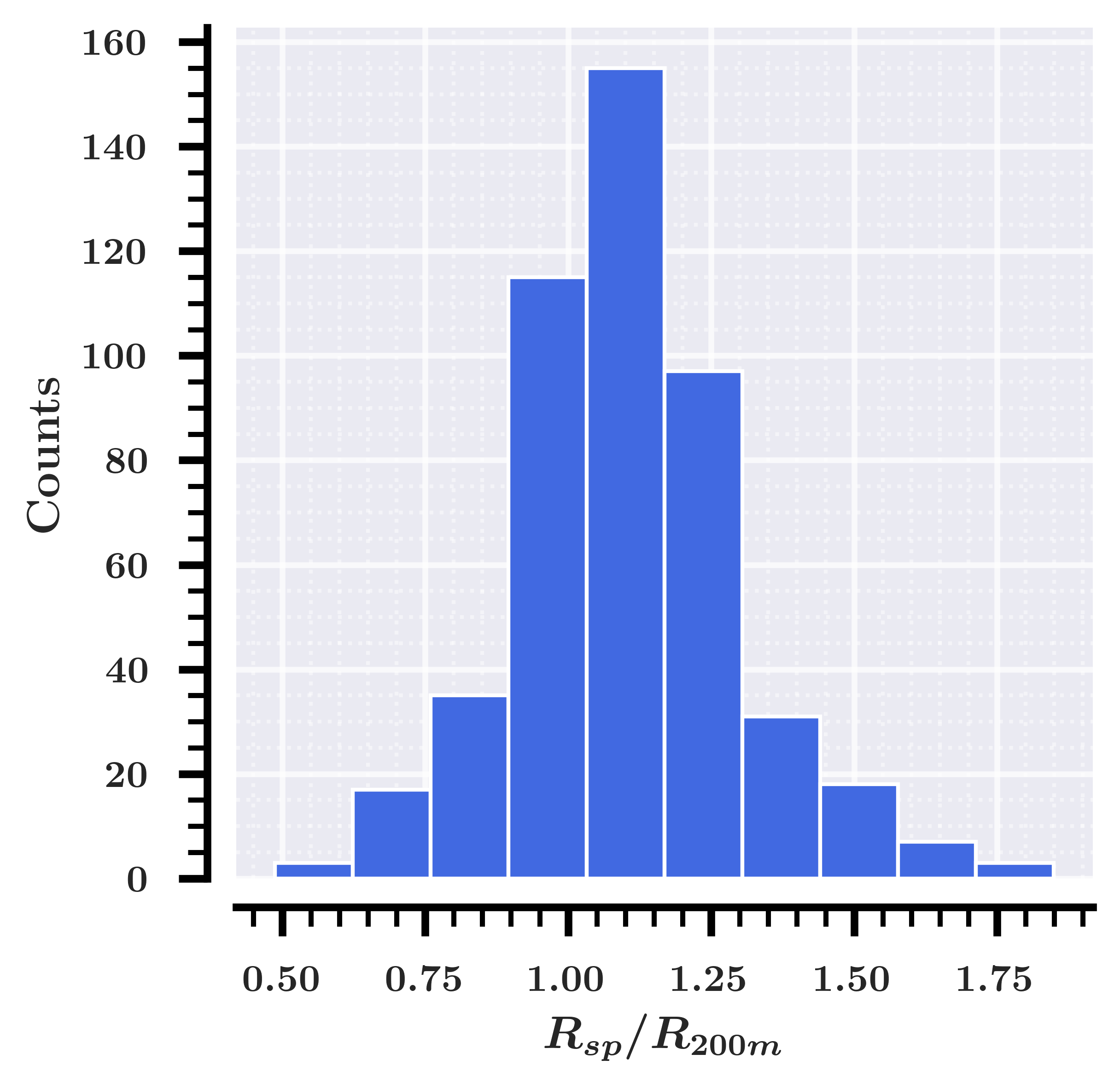}{0.5\textwidth}{(c) Ratio $R_{\mathrm{sp}}/R_{200\mathrm{m}}$.}
\fig{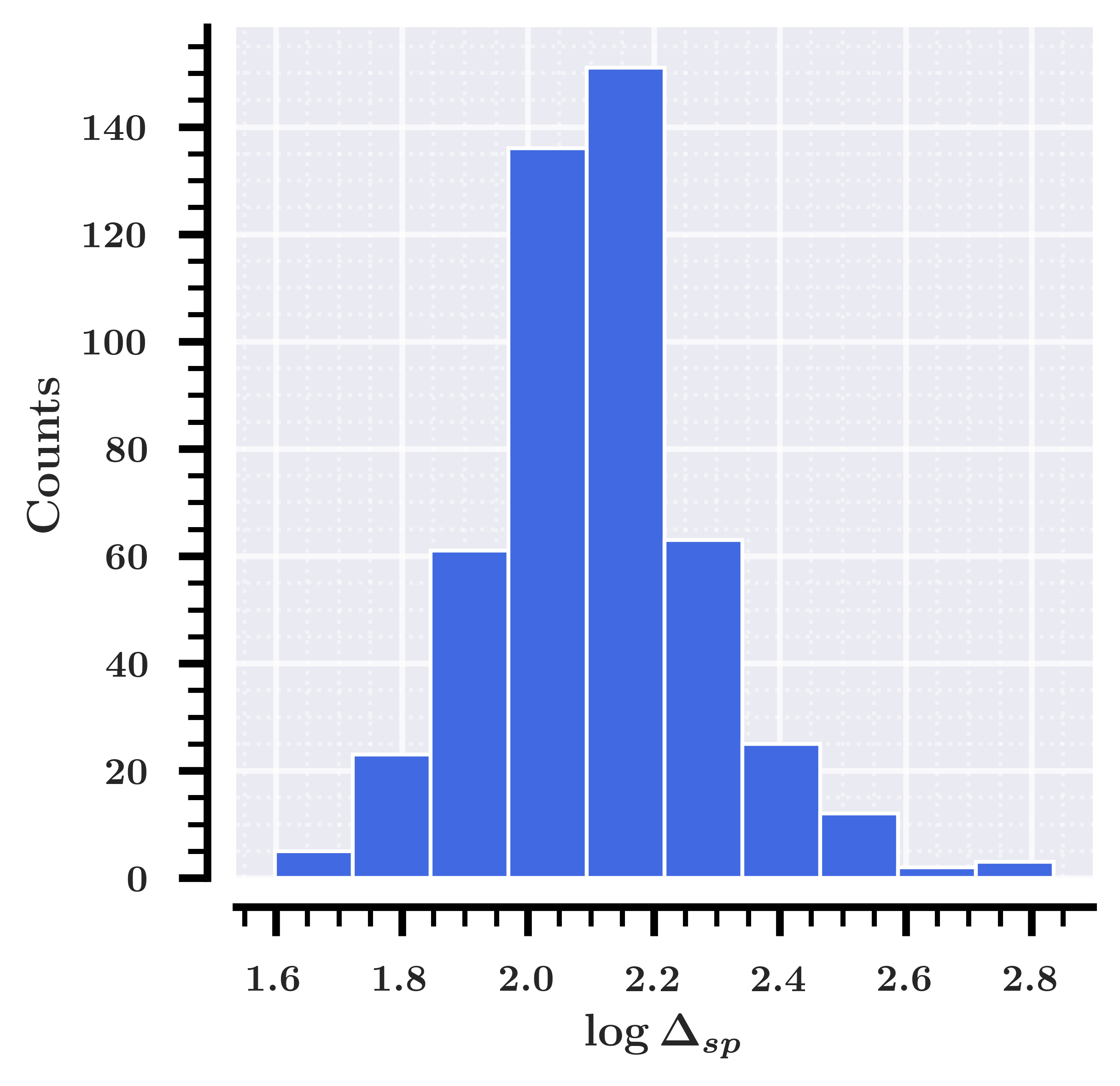}{0.5\textwidth}{(d) Splashback overdensity distribution.}
}
\caption{Distributions of splashback radii, splashback masses, $R_{\mathrm{sp}}/R_{200\mathrm{m}}$ ratio, and enclosed splashback overdensity for the \textsc{CoMaLit} sample.}
\label{fig:comalit_sp_feature}
\end{figure*}

Figure~\ref{fig:comalit_sp_feature} presents the resulting distributions of splashback radius (panel a) and splashback mass (panel b). Panel (c) shows the ratio $R_{\mathrm{sp}}/R_{200\mathrm{m}}$, which is particularly useful for comparison with previous observational studies. For our sample, we find a median value of $R_{\mathrm{sp}}/R_{200\mathrm{m}} \simeq 1.10 \pm 0.01$, fully consistent with earlier measurements \citep[e.g.,][]{more+16, adhikari+21, gabriel-silva+25}. 

Panel (d) displays the distribution of the enclosed splashback overdensity, defined as
\begin{equation}\label{eq:delta_sp}
    \Delta_{\mathrm{sp}} = 
    \frac{M_{\mathrm{sp}}/(4/3\,\pi R_{\mathrm{sp}}^3)}{\rho_m}.
\end{equation}
We find a median overdensity of $\Delta_{\mathrm{sp}} \simeq 127 \pm 2$, close to the characteristic density associated with $R_{100\mathrm{m}}$. Nevertheless, since the splashback radius is not defined with respect to a fixed reference density, a significant intrinsic scatter is expected. This scatter reflects the dependence of $R_{\mathrm{sp}}$ on cluster properties such as mass accretion rate and dynamical state \citep[e.g.,][]{mansfield+17, diemer20}.

With splashback radii and masses measured for the full \textsc{CoMaLit} sample, we recalibrate the $M_{\mathrm{sp}}$--$R_{\mathrm{sp}}$ scaling relation introduced in \citet{gabriel-silva+25}. Motivated by self-similar spherical collapse models \citep{gunn+72}, we assume a power-law dependence including a redshift evolution term. To enable a consistent comparison across redshift and isolate the intrinsic redshift evolution from that induced by cosmic expansion, we rescale mass and radius using the normalized Hubble parameter $E(z)$, motivated by self-similar considerations in which mass and radius are linked through a characteristic density that evolves with the expansion of the Universe \citep[e.g.,][]{bryan+98, white01, hu+03}. This leads to the relation
\begin{equation}\label{eq:mass_fit}
\left(\frac{M_{\mathrm{sp}}}{E(z)^2\,10^{14}M_\odot}\right) =
C \left(\frac{R_{\mathrm{sp}}\,E(z)^{2/3}}{\mathrm{Mpc}}\right)^{A}
(1+z)^{B},
\end{equation}
where $C$ is the normalization, and $A$ and $B$ are the slopes associated with splashback radius and redshift evolution, respectively. Importantly, $E(z)$ is introduced to reduce the redshift dependence driven by cosmic expansion, allowing the parameter $B$ to capture the intrinsic evolution of the relation. It is not intended to correct for pseudo-evolution, as splashback quantities are, by construction, free from this effect.

\begin{figure}
\centering
\includegraphics[width=1.0\linewidth]{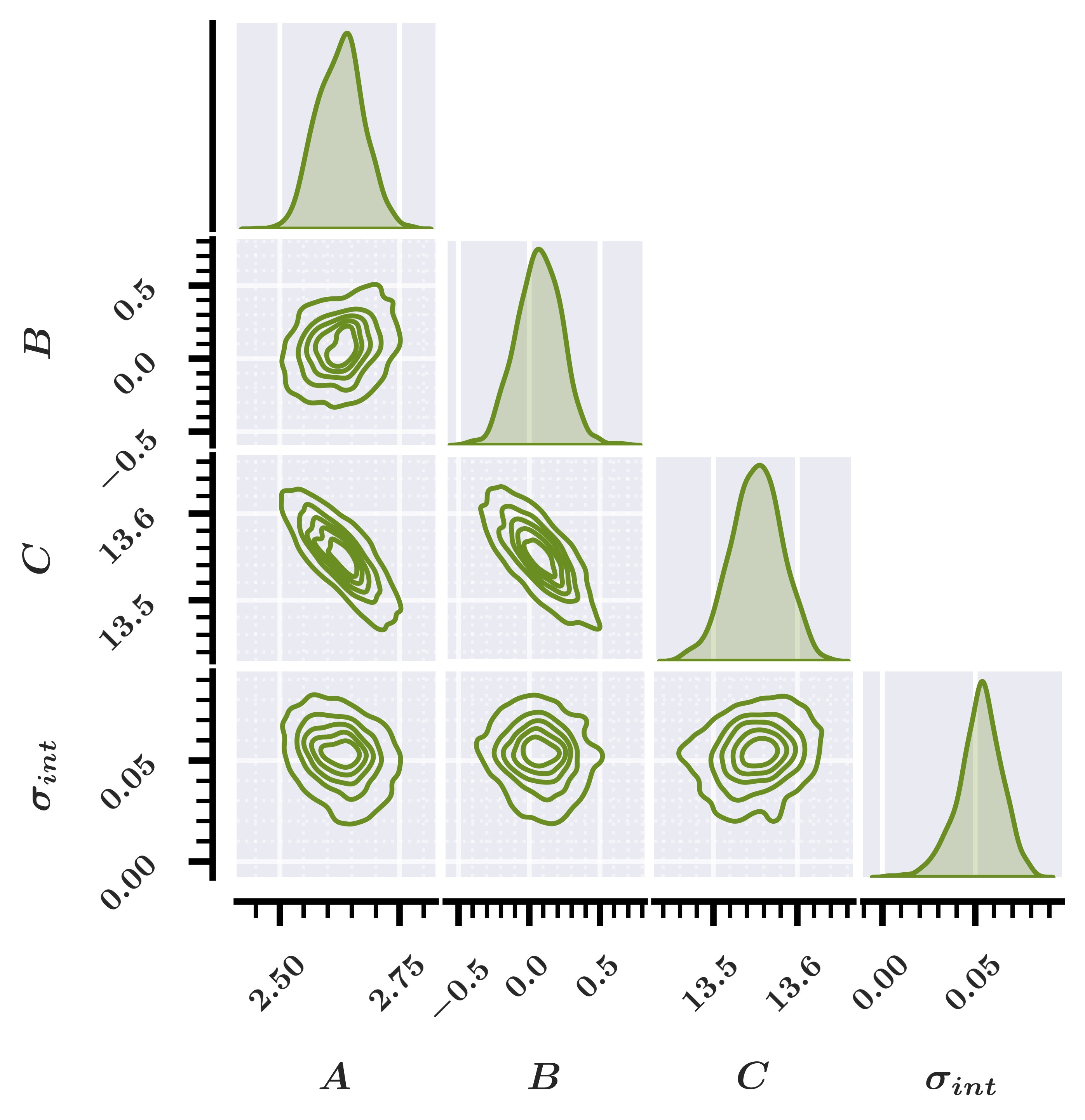}
\caption{Posterior distributions of the $M_{\mathrm{sp}}$--$R_{\mathrm{sp}}$ scaling relation parameters.}
\label{fig:mcmc}
\end{figure}

If clusters shared a characteristic splashback density, one would expect
\begin{equation}
\frac{M_{\mathrm{sp}}}{R_{\mathrm{sp}}^3} = \mathrm{constant},
\end{equation}
implying $A \simeq 3$. Instead, fitting Equation~\ref{eq:mass_fit} with MCMC while including an intrinsic scatter term $\sigma_{\mathrm{int}}$, we obtain $A = 2.63 \pm 0.05$, significantly shallower than the constant-density expectation. We find $C = 0.35 \pm 0.02$, $B = 0.07 \pm 0.15$, and $\sigma_{\mathrm{int}} = 0.053$. The posterior distributions are shown in Figure~\ref{fig:mcmc}.

Although the A slope is lower than 3, this result is not unexpected. Unlike overdensity-defined radii, the splashback radius corresponds to a physical boundary related to the cluster's gravitational potential and mass accretion history, rather than a fixed enclosed density. Its dependence on the mass accretion rate naturally leads to deviations from simple self-similar scaling. In addition, projection effects may play a role, as splashback radii are inferred from two-dimensional galaxy distributions. Nevertheless, compared to our previous work, the larger dataset allows for significantly tighter constraints, and we observe a clear increase in the inferred slope.

\begin{figure}
\centering
\includegraphics[width=1.0\linewidth]{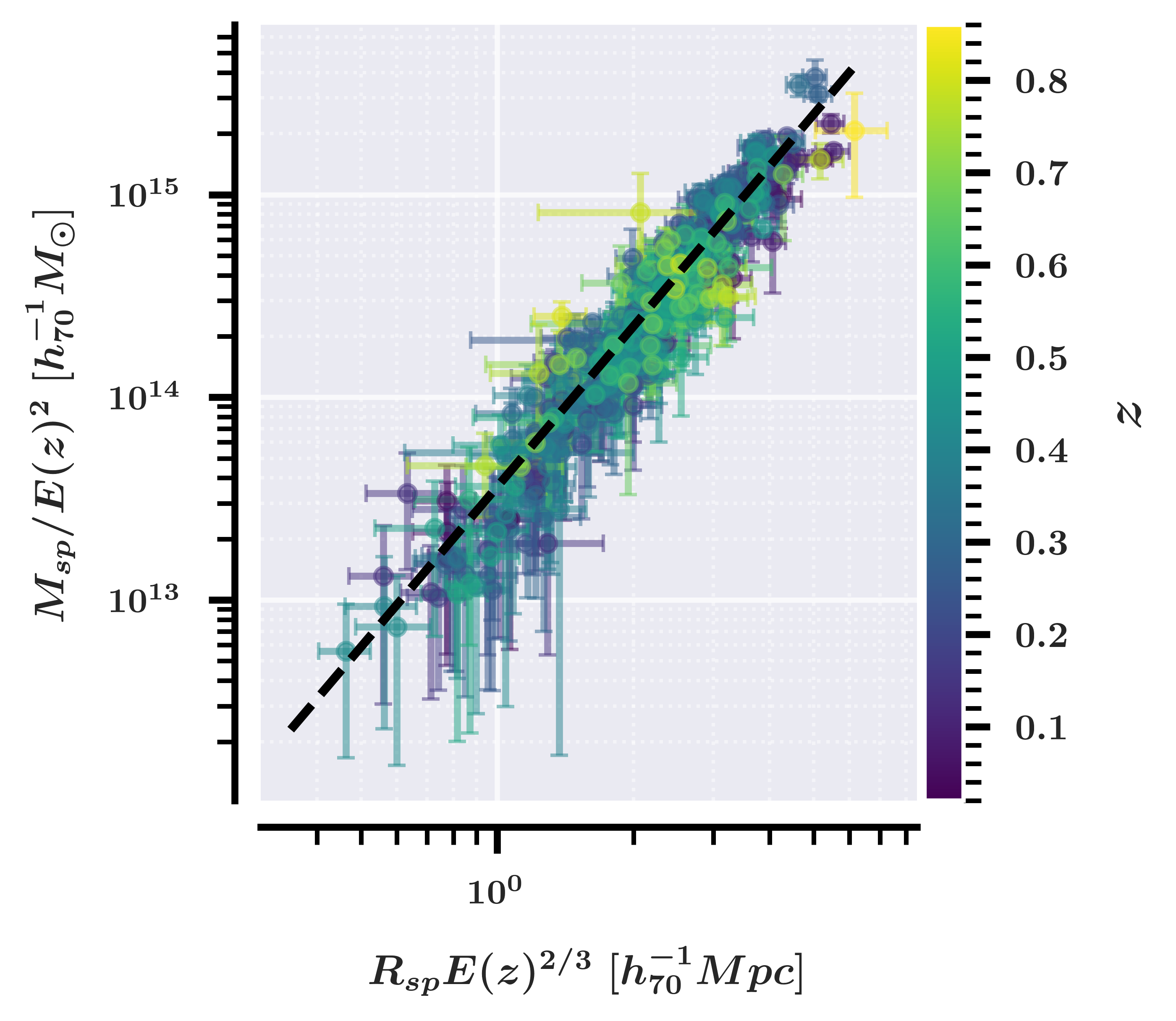}
\caption{$M_{\mathrm{sp}}$--$R_{\mathrm{sp}}$ relation for the \textsc{CoMaLit} sample, color-coded by redshift. The black dotted line shows the best-fit relation.}
\label{fig:mass_fit}
\end{figure}

The redshift slope is fully consistent with zero, indicating no statistically significant evolution for the redshift interval of our sample. This result contrasts with the apparent evolution found in \citet{gabriel-silva+25}, where the limited redshift range prevented a robust determination. Figure~\ref{fig:mass_fit} shows the best-fit $M_{\mathrm{sp}}$--$R_{\mathrm{sp}}$ relation, color-coded by redshift.

Finally, we emphasize that the calibrated scaling relation is competitive for mass estimation purposes. The intrinsic scatter of $\sim0.15$ dex is comparable to that of commonly used mass--richness \citep[e.g.,][]{gonzalez+16, simet+16, murata+18} and mass--luminosity relations \citep[e.g.,][]{popesso+05, popesso+07, mulroy+17}. Since the splashback mass represents a physically motivated halo boundary, this relation offers a promising avenue for estimating cluster masses and sizes in large photometric surveys. In the next subsection, we explore its implications for the construction of a splashback-based cluster mass function.

\subsection{Splashback Mass Function}\label{sec:results:mass_func}

With the calibrated $M_{\mathrm{sp}}$--$R_{\mathrm{sp}}$ relation in hand, we extend our analysis to a much larger volume by considering the SDSS NGC region and the \textsc{redMaPPer} cluster catalog. Applying the same selection criteria and methodology adopted for the \textsc{CoMaLit} sample, we identify 15{,}144 clusters in the redshift range $0.08 < z < 0.6$. For each system, we determine the probabilistic photometric membership, select the optimal adaptive probability cut, and model the cumulative number profile to estimate the splashback radius.

\begin{figure}
\centering
\includegraphics[width=1.0\linewidth]{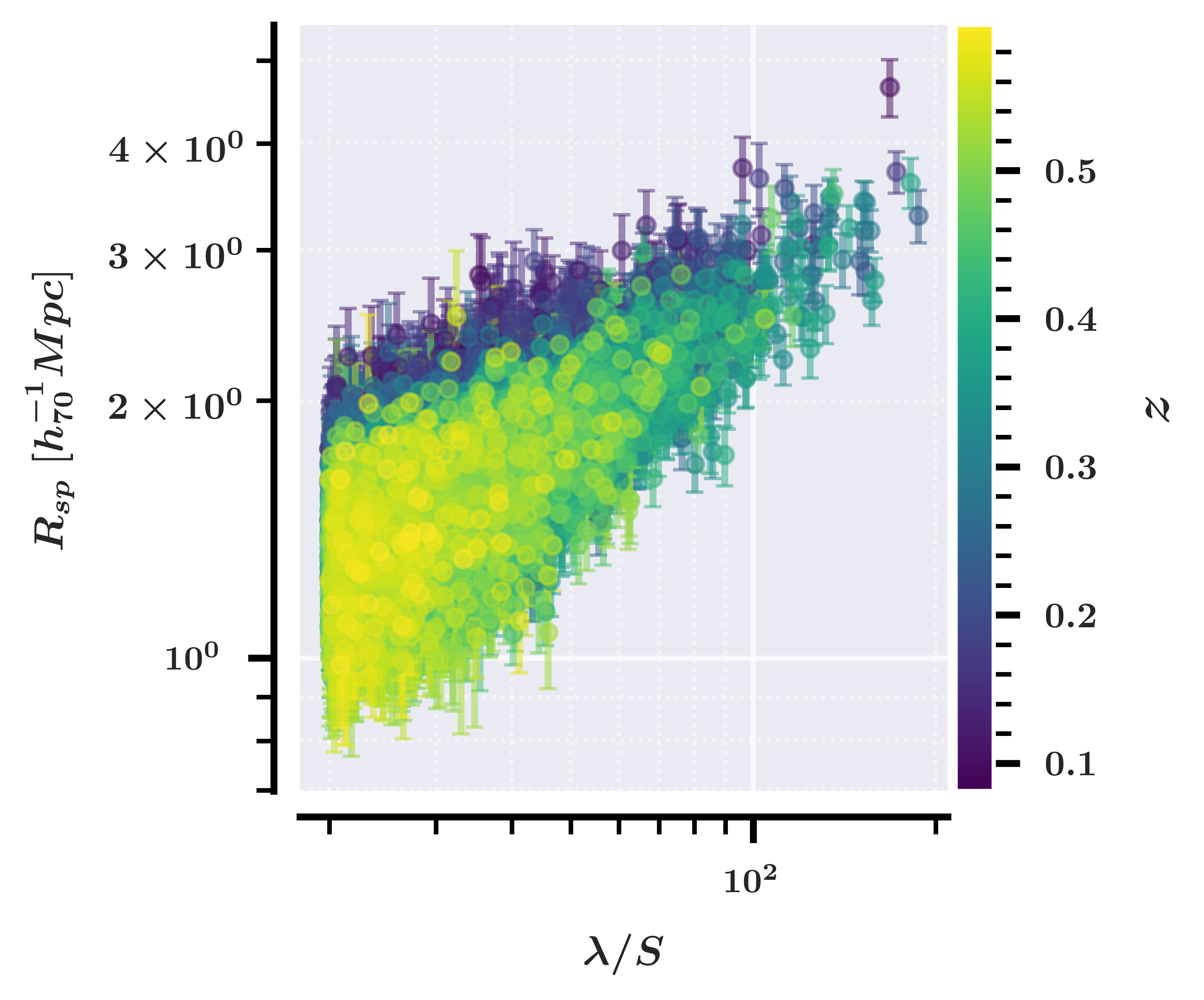}
\caption{Splashback radius as a function of corrected richness for the \textsc{redMaPPer} cluster catalog.}
\label{fig:richness_sp}
\end{figure}

An interesting result is the clear correlation between splashback radius and cluster richness, in agreement with previous findings \citep{rykoff+016}. However, as shown in Figure~\ref{fig:richness_sp}, this relation exhibits substantial scatter. Low-richness clusters tend to have smaller splashback radii on average, but span a wide range of sizes. This behavior is expected and likely driven by observational limitations: in low-richness systems, both the membership assignment and the modeling of the cumulative profile are less robust due to the small number of tracer galaxies.

\begin{figure}
\centering
\includegraphics[width=1.0\linewidth]{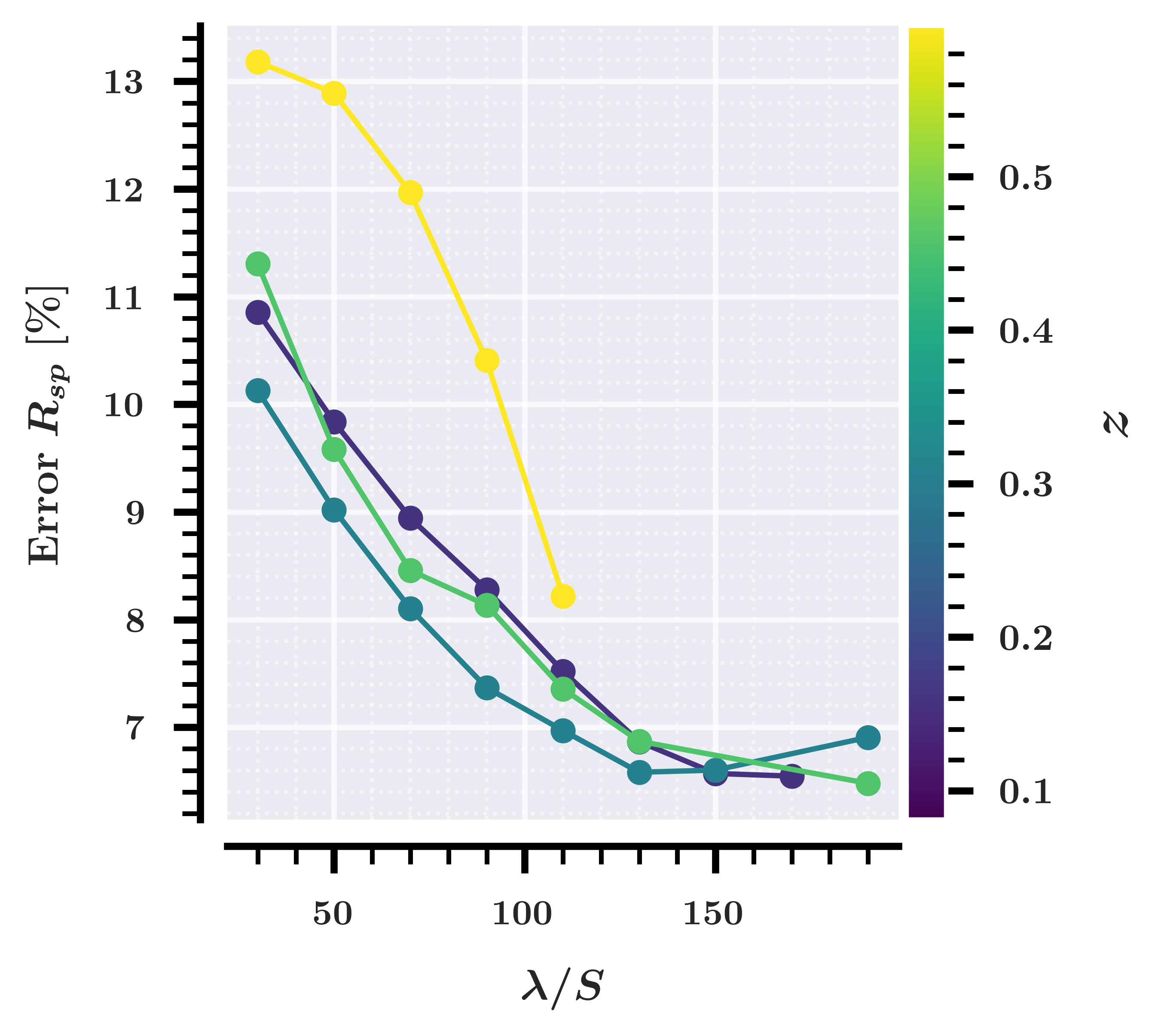}
\caption{Median fractional uncertainty in the splashback radius as a function of cluster richness, color-coded by redshift.}
\label{fig:errors_sp}
\end{figure}

This effect is further illustrated in Figure~\ref{fig:errors_sp}, where we show the median fractional uncertainty in the splashback radius as a function of cluster richness, with redshift encoded in the color scale. Systems with lower richness clearly display larger uncertainties, while higher-richness clusters yield more precise measurements. In addition to richness, the splashback width may also play a role. The width, defined as the full width at half minimum of the logarithmic slope of the density profile \citep{yu+25}, is anti-correlated with halo mass and, by extension, with richness. Although width and uncertainty are distinct quantities, broader splashback features may increase degeneracies among model parameters, indirectly contributing to larger uncertainties in $R_{\mathrm{sp}}$.

\begin{figure*}
\centering
\includegraphics[width=1.0\linewidth]{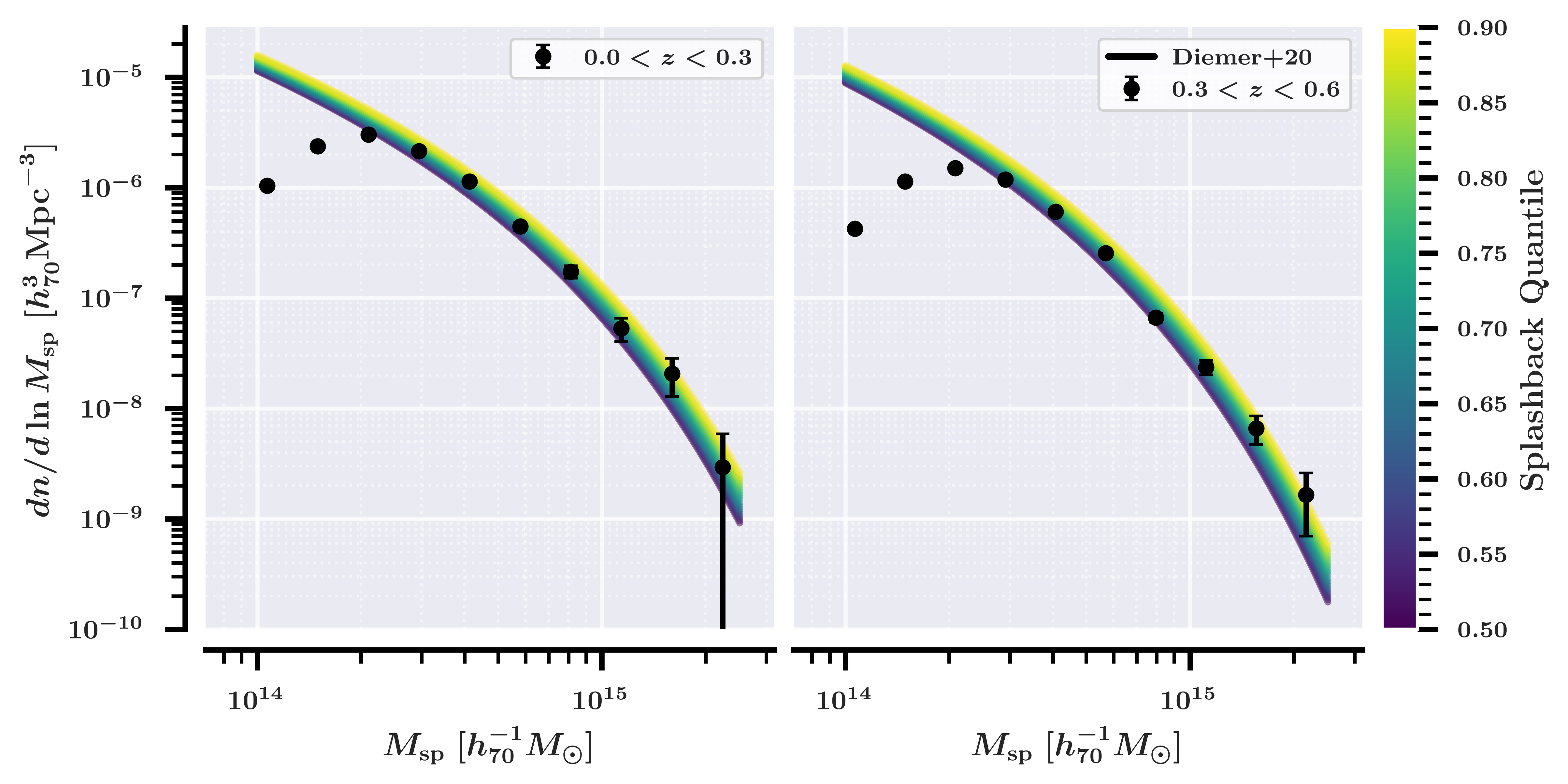}
\caption{Galaxy cluster splashback mass functions for the \textsc{redMaPPer} cluster catalog. Filled lines represent analytical predictions for the splashback mass function calibrated from simulations in \citet{diemer20} and color-coded by the splashabck quantiles.}
\label{fig:mass_func}
\end{figure*}

Despite these effects, the splashback measurements for sufficiently massive systems remain robust. Using the scaling relation in Equation~\ref{eq:mass_fit}, we estimate splashback masses for the \textsc{redMaPPer} clusters and, following the methodology described in Section~\ref{sec:methods:mass_func}, construct a splashback-based cluster mass function. We restrict the analysis to $\log(M_{\mathrm{sp}}/h_{70}^{-1}M_\odot) > 14.0$, where incompleteness effects are expected to be subdominant. The resulting mass functions are shown in Figure~\ref{fig:mass_func} for two redshift bins.

The colored shaded regions in Figure~\ref{fig:mass_func} represent analytical predictions for the splashback mass function calibrated from simulations by \citet{diemer20}, with different colors indicating different splashback quantiles. Although splashback quantiles cannot be directly assigned in observations, one would expect the observed mass function to follow the mean behavior associated with the transition between the one-halo and two-halo regimes. Qualitatively, our measurements are consistent with the $\sim$50--75\% quantile range, particularly at the high-mass end.

A noticeable deviation from the model predictions appears around $M_{\mathrm{sp}} \sim 2 \times 10^{14}\,h_{70}^{-1}M_\odot$, becoming more pronounced in the higher-redshift bin, where the observed mass function declines more rapidly. We interpret this discrepancy primarily as a selection effect rather than a failure of the splashback-based mass estimation. Previous studies comparing \textsc{redMaPPer} clusters with X-ray and SZ-selected samples have shown that the catalog is nearly complete for systems with $T_X \gtrsim 3.5$~keV and $L_X \gtrsim 2 \times 10^{44}\,\mathrm{erg\,s^{-1}}$ at low redshift ($z \lesssim 0.35$), but the completeness drops to $\sim$80--90\% for lower temperatures and luminosities \citep{rykoff+14}. This incompleteness becomes more severe at higher redshifts, as the minimum richness threshold for detection increases \citep{rykoff+016}.

Using standard scaling relations \citep{vikhlinin+09} and conversions between mass definitions, we estimate that these X-ray completeness limits correspond to $M_{200\mathrm{m}} \approx 2 \times 10^{14}\,h_{70}^{-1}M_\odot$. Given that $M_{\mathrm{sp}} \simeq 1.05\,M_{200\mathrm{m}}$ in our sample, this mass scale coincides remarkably well with the point at which our measured mass function begins to deviate from the predictions of \citet{diemer20}. 

\begin{figure}
\centering
\includegraphics[width=\columnwidth]{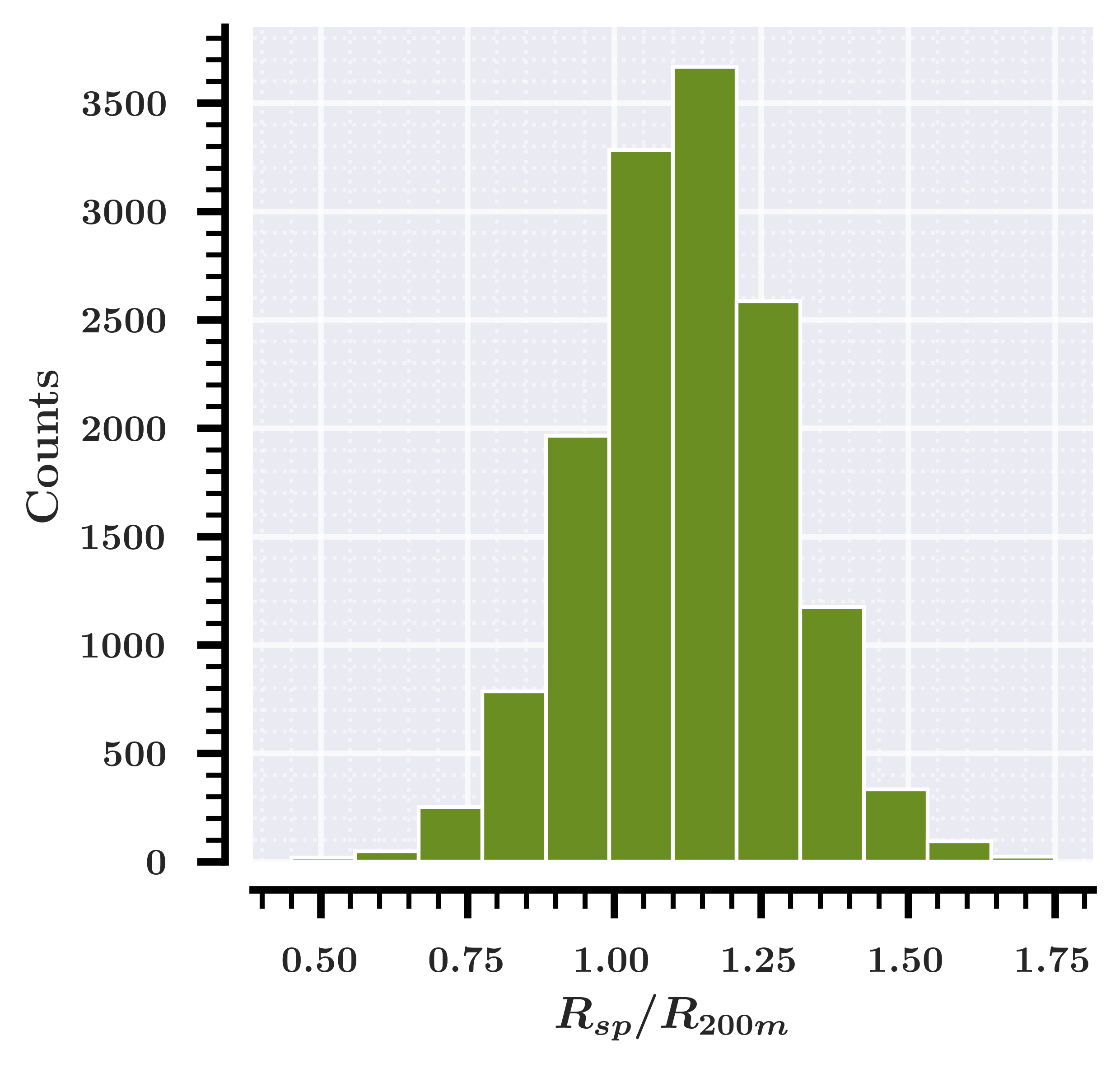}
\caption{Distribution of the ratio $R_{\rm sp}/R_{200\rm m}$ for the redMaPPer cluster catalog.}
\label{fig:redmapper_sp_ratio}
\end{figure}

In addition to completeness effects, it is important to consider potential systematic biases in the splashback radius measurements that can affect mass determinations. Optically selected cluster catalogs such as redMaPPer are known to be biased toward lower splashback radii due to projection and aperture effects \citep[e.g.,][]{more+16,baxter+17,chang+18}. However, we do not observe this behavior in our analysis. In Figure~\ref{fig:redmapper_sp_ratio}, we show the distribution of the ratio $R_{\rm sp}/R_{200\rm m}$ for the redMaPPer catalog. Our estimates cluster around $R_{\rm sp} \sim 1.1\,R_{200\rm m}$, in agreement with values reported for SZ-selected cluster samples \citep[e.g.,][]{shin+19}, as well as expectations from N-body simulations, and in contrast with previous optical studies based on the same catalog that found systematically lower values.

The origin of this difference is not entirely clear, but it may be related to methodological differences. In particular, our estimates are obtained for individual clusters rather than stacked profiles and include an explicit membership selection prior to the splashback determination. As shown in Figure~\ref{fig:sdss_sigmaz}, contamination, mainly driven by photometric redshift uncertainties, can significantly bias $R_{\rm sp}$ toward lower values if not properly controlled. In this sense, the adopted membership procedure likely helps mitigate this effect, leading to splashback radii that do not exhibit the expected bias toward smaller values. Finally, since the $M_{\rm sp}$--$R_{\rm sp}$ relation is calibrated using a sample with mixed selection \citep{sereno15}, we do not expect any residual systematic effects in $R_{\rm sp}$ to significantly impact the resulting mass function.

Overall, the agreement at the high-mass end is striking, demonstrating that our methodology is capable of recovering the expected cosmological mass distribution using splashback masses derived solely from photometric data. This result highlights the potential of splashback-based observables as a physically motivated and powerful tool for cluster studies in current and future large photometric surveys.

\section{Conclusions}\label{sec:conclusion}

Galaxy clusters provide a powerful laboratory for testing models of structure formation and cosmology. In particular, accurate measurements of cluster masses are essential for discriminating between cosmological scenarios. However, traditional mass and radius definitions based on spherical overdensities are affected by pseudo-evolution \citep{diemer+13}, which can bias interpretations of halo growth. The splashback feature offers a physically motivated alternative, defining the outer boundary of dark matter halos as the transition between virialized and infalling material. By construction, the splashback radius does not suffer from pseudo-evolution, making it a natural scale for defining cluster size and mass and a promising basis for constructing cluster mass functions.

In this work, we develop and apply a probabilistic photometric membership method, based on SDSS photometric redshifts, following the framework introduced by L\"osch et al. (in preparation). Using these memberships, we model the cumulative number density profiles of individual galaxy clusters from the \textsc{CoMaLit} catalog to estimate their splashback radii as the minima of the logarithmic slope of the density profile. We further infer splashback masses by extending weak-lensing $M_{200\mathrm{c}}$ measurements with analytical halo models. Spectroscopic data are used exclusively for validation purposes, ensuring that the method remains fully applicable to purely photometric samples. This work extends the analysis of \citet{gabriel-silva+25} to a significantly larger dataset and a broader redshift range.

We summarize our main results as follows:

\begin{itemize}

\item We show that our probabilistic photometric membership method performs robustly when compared to spectroscopic memberships. Combining redshift and radial probability components, the method achieves typical completeness and purity of $\sim$70\% within $3R_{200\mathrm{m}}$ for a probability threshold of 0.5, increasing to $\sim$90\% completeness and $\sim$85\% purity within $R_{200\mathrm{m}}$ (Figure~\ref{fig:comp_purity}). These values are comparable to previous observational studies adopting similar approaches \citep[e.g.,][]{castigani+16}. While recovering individual galaxy memberships is not our primary goal, this level of performance is sufficient to accurately recover the cluster density profiles required for splashback measurements.

\item Rather than adopting a fixed probability threshold for all systems, we demonstrate that an adaptive probability cut, chosen to maximize the detection significance between the cluster core ($R_\mathrm{core} = R_{200\mathrm{m}}/2$) and outskirts ($R_{200\mathrm{m}} < R < 3R_{200\mathrm{m}}$), provides a more optimal definition of cluster membership. Using the formulation of \citet{lima83}, we find that the optimal cut varies from cluster to cluster, typically peaking around $P_{\mathrm{cut}} \sim 0.6$ for nearby, rich systems, but reaching values as low as $\sim$0.2--0.3 for higher-redshift or poorer clusters (Figure~\ref{fig:SN_Pcut}). This adaptive approach yields splashback radius estimates that are more consistent with spectroscopic measurements from \citet{gabriel-silva+25} (Figure~\ref{fig:sdss_sigmaz}).

\item From the resulting memberships, we model the cumulative number profiles of 499 clusters in the \textsc{CoMaLit} sample using MCMC techniques. Splashback radii are identified as the minima in the logarithmic slope of the density profile, and splashback masses are estimated accordingly. The distributions of $R_{\mathrm{sp}}$, $M_{\mathrm{sp}}$, and especially the ratio $R_{\mathrm{sp}}/R_{200\mathrm{m}}$ are consistent with previous observational studies, with a median ratio of $R_{\mathrm{sp}}/R_{200\mathrm{m}} \simeq 1.10 \pm 0.01$ \citep[e.g.,][]{more+16, adhikari+21, gabriel-silva+25}. We also find a broad distribution of splashback overdensities, as expected for a physically defined boundary not tied to a reference density, with a median value of $\Delta_{\mathrm{sp}} = 127 \pm 2$ (Figure \ref{fig:comalit_sp_feature}).

\item Using this expanded dataset, we recalibrate the $M_{\mathrm{sp}}$--$R_{\mathrm{sp}}$ scaling relation originally presented by \citet{gabriel-silva+25} (Figure~\ref{fig:mass_fit}). By substantially increasing both the sample size and the redshift leverage ($0.01 < z < 0.8$), and after explicitly removing the cosmological dependence through the $E(z)$ rescaling, we find no statistically significant redshift evolution in the relation, with the best-fit redshift slope fully consistent with zero. The best-fit slope relating mass and radius is significantly shallower than the self-similar expectation of $A=3$, consistent with our previous findings. This deviation likely reflects a combination of projection effects and the fact that the splashback radius is not defined by a fixed overdensity, but instead depends on the cluster dynamical state and mass accretion rate \citep[e.g.,][]{mansfield+17, diemer20}. Despite this, the intrinsic scatter of the relation remains low ($\sim$0.15 dex), comparable to other widely used mass--observable relations such as mass--richness and mass--luminosity \citep[e.g.,][]{gonzalez+16, mulroy+17, murata+18}.

\item Finally, we apply our calibrated scaling relation to the \textsc{redMaPPer} catalog in the SDSS North Galactic Cap, analyzing over 15{,}000 clusters in the redshift range $0.08 < z < 0.6$. We find a clear correlation between splashback radius and cluster richness (Figure \ref{fig:richness_sp}), albeit with significant scatter, particularly for low-richness systems where uncertainties in membership and profile modeling are larger (Figure \ref{fig:errors_sp}). Using the inferred splashback masses, we construct cluster mass functions in two redshift bins and find excellent agreement with the analytical predictions of \citet{diemer20} at the high-mass end (Figure \ref{fig:mass_func}). Deviations at lower masses, around $M_{\mathrm{sp}} \sim 2 \times 10^{14}\,h_{70}^{-1}M_\odot$, are consistent with known completeness limitations of the \textsc{redMaPPer} catalog \citep{rykoff+14}, rather than shortcomings of the splashback-based mass estimation.

\end{itemize}

Overall, our results demonstrate that the splashback feature is a robust and physically meaningful observable that can be measured for individual clusters using photometric data alone. The ability to estimate both cluster size and mass from splashback measurements opens new avenues for cosmological analyses in large photometric surveys. The close agreement between our observational mass functions and theoretical expectations further supports the use of splashback-based quantities as reliable probes of halo structure and evolution.

Future work will extend this analysis to larger survey areas and deeper photometric datasets, improving statistical precision and enabling quantitative comparisons with theoretical splashback mass function models. In particular, forthcoming surveys such as the Rubin Observatory Legacy Survey of Space and Time \citep{ivezic+19} will provide an ideal dataset to exploit the full potential of splashback-based cluster studies.


\section*{Acknowledgements}
\begin{acknowledgments}
LG-S was financed by the São Paulo Research Foundation (FAPESP), Brasil (2024/03449-3). LSJ ackowlegdes the support from CNPq (308994/2021-3) and FAPESP (2011/51680-6).
\end{acknowledgments}

\vspace{5mm}
\facilities{Sloan}

\software{Astropy \citep{astropy:2013, astropy:2018, astropy:2022},
          COLOSSUS \citep{colossus},
          emcee \citep{emcee},
          NumPy \citep{numpy},
          SciPy \citep{scipy}}
          


\bibliography{sample631}{}
\bibliographystyle{aasjournal}



\end{document}